\documentclass{article} % For LaTeX2e
\usepackage{iclr2025_conference,times}

\usepackage{amsmath,amsfonts,bm}

\def\eqref#1{equation~\ref{#1}}
\def\1{\bm{1}}

\DeclareMathAlphabet{\mathsfit}{\encodingdefault}{\sfdefault}{m}{sl}
\SetMathAlphabet{\mathsfit}{bold}{\encodingdefault}{\sfdefault}{bx}{n}

\usepackage{arydshln}
\usepackage{hyperref}
\usepackage{url}
\usepackage{xcolor}
\usepackage{graphicx}
\usepackage{colortbl}
\usepackage{booktabs}
\usepackage{multirow}
\usepackage{authblk}

\title{Making Text Embedders Few-Shot Learners}

\author{Chaofan Li$^{1,2}$\thanks{Co-first authors}, MingHao Qin$^{1,3,*}$, Shitao Xiao$^{1}$, Jianlyu Chen$^{1,4}$, Kun Luo$^{1,3}$, Yingxia Shao$^{2}$, Defu Lian$^{4}$, Zheng Liu$^{1}$\thanks{Corresponding author} \\
1: Beijing Academy of Artificial Intelligence\\
2: Beijing University of Posts and Telecommunications\\
3: Chinese Academy of Sciences\\
4: University of Science and Technology of China\\
\texttt{\{cfli, shaoyx\}@bupt.edu.cn} \quad
\texttt{qinminghao24@ia.ac.cn} \\
\texttt{stxiao@baai.ac.cn} \quad
\texttt{chenjianlv@mail.ustc.edu.cn} \\
\texttt{liandefu@ustc.edu.cn} \quad
\texttt{\{luokun695, zhengliu1026\}@gmail.com}
}

\iclrfinalcopy % Uncomment for camera-ready version, but NOT for submission.

\begin{document}

\maketitle

\begin{abstract}
% In this paper, we rethink the application of LLM in dense retrieval.
% How to use LLM to generate embedding
% Decoder-only large language models (LLMs) have demonstrated remarkable efficacy as embedding models. Recent research has increasingly focused on modifying the architecture of these models, particularly by shifting the attention mechanism towards a bidirectional format. Contrary to this trend, our work rethinks the application of LLMs in dense retrieval tasks, advocating for the preservation of the original architecture. Considering that LLMs develop robust in-context learning (ICL) capabilities following their pre-training phase, we propose to integrate this ability into dense retrieval tasks. Specifically, the queries are expanded and augmented with task-relevant few-shot examples to enhance the semantic representation of the query's text embeddings, which is expected to improve the model's domain-specific performance and its generalization abilities. Our empirical results validate that utilizing the ICL training strategy on the same training dataset leads to a significant improvement in model performance. Moreover, when trained with full dataset, our model exhibits state-of-the-art results on both the Massive Text Embeddings (MTEB) Benchmark and AIR-Bench benchmark.

% kunluo
% 目前embedder模型广泛探索并使用LLM作为基座
% 他们有些更改注意力机制，有些使用复杂的多阶段训练
% 然而，他们在OOD任务上泛化能力、对于有不同检索意图任务的指令遵循能力依然受限
% 介绍LLM本身具有ICL能力能够进行任务泛化
% 我们认为应该尝试激发LLM在处理embedding任务时的ICL能力，从而构建泛化能力更强、更为通用的embedder
Large language models (LLMs) with decoder-only architectures demonstrate remarkable in-context learning (ICL) capabilities. This feature enables them to effectively handle both familiar and novel tasks by utilizing examples provided within their input context. Recognizing the potential of this capability, we propose leveraging the ICL feature in LLMs to enhance the process of text embedding generation.
To this end, we introduce a novel model \textbf{bge-en-icl}, which employs few-shot examples to produce high-quality text embeddings. Our approach integrates task-related examples directly into the query side, resulting in significant improvements across various tasks.
Additionally, we have investigated how to effectively utilize LLMs as embedding models, including various attention mechanisms, pooling methods, etc. Our findings suggest that retaining the original framework often yields the best results, underscoring that simplicity is best. Experimental results on the MTEB and AIR-Bench benchmarks demonstrate that our approach sets new state-of-the-art (SOTA) performance.
Our model, code and dataset are freely available at \href{https://github.com/FlagOpen/FlagEmbedding}{https://github.com/FlagOpen/FlagEmbedding}.
\end{abstract}

\section{Introduction}

Text embeddings are vector representations that capture the semantic and contextual meaning of natural language text. They play a pivotal role in natural language processing (NLP) tasks, facilitating a wide range of applications such as information retrieval, text classification, item recommendation, and question answering \citep{karpukhin2020dense, xiong2020approximate, lu2020twinbert}. Pre-trained bidirectional encoder and encoder-decoder architectures have been widely adopted as backbone models for embedding model, owing to their effectiveness in producing high-quality vector embeddings for text thanks to their extensive pre-training \citep{xiao2022retromae, gao2021simcse}.

Recent advancements in LLMs have significantly shifted the focus towards embedding models that rely primarily on decoder-only architectures \citep{ma2023fine, li2024llama2vec, wang2023improving}. These LLM-based embedding models have demonstrated remarkable improvements in in-domain accuracy and generalization, particularly when trained using supervised learning approaches \citep{wang2023improving}. However, despite these advances, embedding models still struggle to follow unseen task instructions and execute complex retrieval tasks \citet{su2024bright, weller2024followir}. This limitation stems from a mismatch between the relatively narrow range of instructions encountered during training and the broader variety of real-world text embedding tasks.

In-context learning (ICL) is a core capability of LLMs, enabling them to incorporate task-specific examples directly into input prompts to generate desired outputs \citep{radford2019language, brown2020language, gao2020making}. The scope of ICL extends beyond tasks seen during training; it enables LLMs to generalize to new and complex tasks by learning patterns from the provided examples. This allows LLMs to adapt dynamically to novel tasks without additional training, making them highly applicable to large-scale, real-world scenarios \citep{wei2022chain, yao2022react, dong2022survey}.

Recognizing the robust ICL abilities of LLMs, in this study, we propose to generate more adaptable text embeddings with ICL strategy. Specifically, we guide the model by including task-specific examples directly within the query prompt. By doing so, we leverage the ICL capabilities of LLMs to produce embeddings that are not only more relevant to the specific domain but also more generalizable across various contexts.

Moreover, LLMs are predominantly utilized for text generation tasks, and adapting them for embedding representation tasks requires specific fine-tuning strategies. Recent studies have introduced various approaches, including the generation of high-quality training data through LLMs \citep{wang2023improving}, modifications to attention mechanisms, and changes in pooling methods \citep{ma2023fine, li2024llama2vec}.
Following previous works \citep{muennighoff2024generative, behnamghader2024llm2vec}, we investigate how to effectively utilize LLMs as embedding models by modifying various architectures, e.g., bidirectional attention, meaning pooling, etc.
Our experimental findings indicate that in the ICL scenario, making complex modifications to the models does not lead to significant improvements. Surprisingly, the best results are obtained using the original, unmodified architecture.
By employing only the ICL strategy, our model \textbf{bge-en-icl} achieves state-of-the-art (SOTA) results on both the MTEB and AIR-Bench benchmarks. We have also released a multi-language embedding model \textbf{bge-multilingual-gemma2} and a lightweight reranker \textbf{bge-reranker-v2.5-gemma2-lightweight}. The lightweight reranker also serves as the teacher model for training embedding models through distillation. Further details are provided in Appendices \ref{multilingual} and \ref{reranker}.

In summary, the key contributions of our work are as follows:
\begin{itemize}
    \item We propose \textbf{bge-en-icl}, which incorporate few-shot examples into the query side to enhance the query embeddings. This integration leverages the in-context learning (ICL) capabilities of large language models (LLMs) in text embedding tasks.
    \item We rethink and explore how to effectively utilize LLMs as embedding models by evaluating various attention mechanisms, pooling methods, and the incorporation of passage prompts.  Our findings highlight that simplicity is best; simply combining ICL capabilities with embedding models can achieve excellent performance.
    % \item We rethink the application of embedding models by evaluating various attention mechanisms, pooling methods, and the incorporation of passage prompts.  Our findings highlight an effective approach for utilizing LLMs as embedding models, specifically recommending the use of causal attention, last token pooling, and the exclusion of passage prompts.
    \item In contrast to other leading models on the MTEB benchmark, we provide open access to our model checkpoint, dataset, and training scripts.
\end{itemize}

\section{Related Work}

% 1. Large Language Models as Foundations for Next-Gen Dense Retrieval: A Comprehensive Empirical Assessment 这种related work 写法是以按点串联的方式进行介绍的，直到引出一个面
% 2. llara 的写法则是从一个面开始的，从中找到一个点，然后根据这个点，串联到下一个面。
% 对于这篇文章来说，我觉得按第二个来写比较好，第一个有点啰嗦。

% 撰写思路：
% 第一个面：什么是 dense retrieval， 他为什么重要
% 点：预训练模型是影响retriever的关键因素，，所以现在大家都慢慢地倾向于用越来越大的模型
% 第二个面：各种流派的 LLM-Based retriever 
% 点：他们没有很好的利用llm最为重要的特性 icl ability

Text embedding is a critical research direction in the field of information retrieval, with wide-ranging applications including web search, question answering, and dialogue systems. The fundamental principle involves encoding both queries and documents into embedding vectors within the same latent space. By calculating similarity scores between these vectors, effective retrieval is achieved.
In recent years, numerous studies have leveraged pre-trained language models such as BERT \citep{devlin2018bert}, T5 \citep{raffel2020exploring}, and RoBERTa \citep{liu2019roberta} as the backbone for embedding models. These models have consistently demonstrated superior performance compared to traditional sparse retrieval methods.

The capability of the backbone is a crucial determinant in the effectiveness of retrieval systems. \citep{luo2024large} have demonstrated that performance improves with increased scale and extensive pre-training. Currently, numerous studies have explored the effectiveness of utilizing LLMs as backbone encoders for text embedding tasks.

Repllama \citep{ma2023fine} fine-tuned Llama-2 to serve as both a dense retriever and a reranker, demonstrating the effectiveness of applying large language models (LLMs) in text embedding tasks. To further align LLMs with text embedding tasks, Llama2Vec \citep{li2024llama2vec} introduced two pretraining tasks specifically designed to enhance the model's performance in such tasks, which led to significant improvements on the BEIR benchmark. E5-mistral and Gecko \citep{wang2023improving, lee2024gecko} advanced the training of LLM-based embedding models through the use of synthetic data, markedly boosting their performance across a diverse range of retrieval and non-retrieval tasks.
NV-Embed \citep{lee2024nv} innovatively proposed a latent attention layer to replace conventional pooling methods and implemented a two-stage training strategy to address the challenge of false negatives in non-retrieval tasks. This model has shown strong performance in both retrieval and non-retrieval domains. Additionally, GRIT \citep{muennighoff2024generative} successfully integrated text embedding and generation within a single LLM, achieving performance levels on par with specialized models focused solely on either embedding or generation.
In the exploration of LLMs as embedding models from an unsupervised perspective, LLM2Vec \citep{behnamghader2024llm2vec} presented a novel unsupervised method to transform decoder-only LLMs into embedding models. This approach demonstrated significant potential for modifying LLM backbone encoders to perform retrieval without any supervision. Similarly, PromptReps \citep{zhuang2024promptreps} leveraged chat-based LLMs aligned with human preferences to generate high-quality dense representations in an unsupervised manner.

The LLM-based embedding models mentioned above exhibit commendable performance across both retrieval and non-retrieval tasks. However, much of the existing work has disproportionately focused on altering model architectures, thereby neglecting the intrinsic capabilities of LLMs. Even models like GritLM, which integrate generation and embedding functionalities, fail to fully exploit the potential ICL capabilities of LLMs within the embedding process. By leveraging the innate ICL capabilities of LLMs, embedding models can be more versatile and adapt to diverse scenarios without necessitating additional fine-tuning.
Our model not only achieves SOTA results on the MTEB and AIR-Bench benchmarks but also effectively utilizes the inherent strengths of LLMs across tasks.

\section{Methology}

\begin{figure}[t]
\centering
\centering
\vspace{-14pt}
\includegraphics[width=0.97\linewidth,  height=0.712\linewidth]{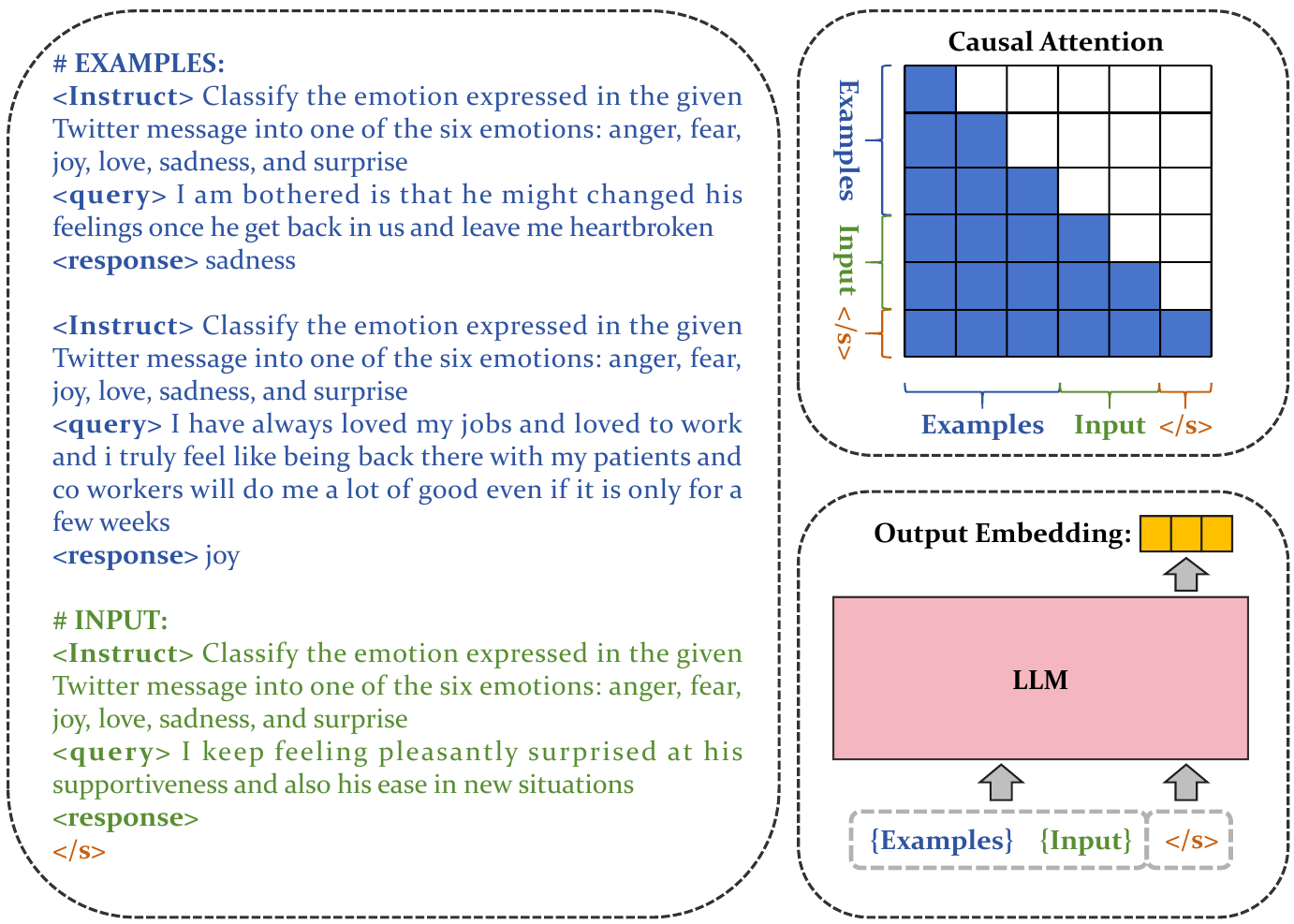}
\vspace{-6pt}
\caption{The architecture of the ICL-based model.}
\vspace{-10pt}
\label{fig:new}
\end{figure}

\subsection{In-Context Learning for Embedding Models}
% few-shot embedding with ICL
% - general purpose (from ICL in LM to ICL in embedding)
% - introduce example template and query template
% - detailed training (and evaluation) method

Previous embedding models often involve directly inputting the query into the model to generate target embeddings. However, this method struggles to handle tasks with different intents, limiting the model's adaptability and generalization capabilities. To address this, researchers have introduced task instructions \citep{su2022one} appended to queries, enabling a single embedding model to generalize across tasks in various domains by altering the instructions.

Despite these advances, studies such as \citet{su2024bright, weller2024followir} reveal that embedding models have a limited ability to follow unseen embedding task instructions and conduct complex retrieval tasks. This limitation arises from a gap between the limited diversity of instructions seen during training and the vast range of real-world scenarios. Inspired by the ability of LLMs to generalize to unseen tasks through in-context learning (ICL), we explore whether embedding models can be enhanced by leveraging ICL, thereby significantly improving their generalization and versatility across diverse embedding tasks with various user intents.

In this work, we demonstrate the potential of embedding models to benefit from ICL through few-shot contrastive training. Consider a query-passage pair $(\mathrm{q}_i, \mathrm{p}_i)$ in an embedding task. We first construct an example template as follows: \begin{equation}
    \langle \text{Instruct} \rangle \hspace{0.2cm} \{\text{task\_definition}\} \newline
    \hspace{0.2cm} \langle \text{query} \rangle \hspace{0.2cm} \{q_i\} \newline
    \hspace{0.2cm} \langle \text{response} \rangle \hspace{0.2cm} \{p_i\}
\end{equation}

Here, "\textit{task\_definition}" represents the description of the specific embedding task. This example template is applied to new input queries for each embedding task (Figure \ref{fig:new}). For a relevant query-passage pair $(\mathrm{q}^+, \mathrm{p}^+)$, the modified query $\mathrm{q}^+_{\mathrm{exp}}$ is constructed as follows: 
\begin{equation}
    \{\text{example 1}\} \hspace{0.1cm} ... \hspace{0.1cm} \{\text{example n}\} \hspace{0.2cm}
    \langle \text{Instruct} \rangle \hspace{0.2cm} \{\text{task\_definition}\} \newline
    \hspace{0.2cm} \langle \text{query} \rangle \hspace{0.2cm} \{q^+\} \newline
    \hspace{0.2cm} \langle \text{response} \rangle
\end{equation}

All modified queries and passages in the corpus are encoded using the same LLM to obtain their embedding representations. Specifically, we append an [EOS] token to the end of the input modified queries and passages, feeding them into the LLM to obtain embeddings $(\mathrm{h}_{\mathrm{q}^+_{\mathrm{exp}}}, \mathrm{h}_{\mathrm{p^+}})$ by extracting the final layer's [EOS] vector. We employ the standard InfoNCE \citep{izacard2021unsupervised} loss function $\mathrm{L}$, utilizing both in-batch negatives and hard negatives for training: \
\begin{equation}
    \mathrm{L} = - \log \frac{\exp(\mathrm{s}(\mathrm{q}^+_{\mathrm{exp}}, \mathrm{p}_i^+))}{
    \exp(\mathrm{s}(\mathrm{q}^+_{\mathrm{exp}}, \mathrm{p}_i^+)) + \sum\limits_{j} \exp(\mathrm{s}(\mathrm{q}^+_{\mathrm{exp}}, \mathrm{p}_j^-))}
    \label{eq:loss_function}
\end{equation}

$\mathrm{p}_j^-$ denotes the set of negative passages, and $\mathrm{s}(\mathrm{q}, \mathrm{p})$ is the scoring function between the query and passage. In this work, we adopt a temperature-scaled cosine similarity function defined as: \begin{equation} \mathrm{s}(\mathrm{q}, \mathrm{p}) = \frac{1}{\tau} \cos(\mathrm{h}_q, \mathrm{h}_p) \end{equation} where $\tau$ is a temperature hyperparameter, which is fixed at 0.02 during training.

% \begin{figure}[t]
% \centering
% \centering
% \vspace{-18pt}
% \includegraphics[width=0.97\linewidth,  height=0.412\linewidth]{figures/model.png}
% \vspace{-6pt}
% \caption{Model architecture}
% \vspace{-10pt}
% \label{fig:model}
% \end{figure}

\subsection{Representation Method}
% Minghao
% 简单说了一下 使用了单向？ 动机是 双向的没有意义？
% 先介绍一下单双向注意力的主要区别，他们分别用于xx模型，然后讲目前许多工作认为讲llm的单向改为双向更有利于embedding场景，如nv embed， gte gritlm， 最后说，我们认为这样丧失了llm的， next token能力？要不要说到icl上？

% Chaofan
% 双向修改了原模型的注意力机制，将decoder架构更改成了类似encoder的架构，下游任务与预训练目标差距过大

% Kunluo
% 在该点同时介绍attention方法和pooling方法

The attention mechanism in LLM-based embedding models is typically unidirectional, aligned with the next-token prediction task fundamental to their pre-training \citep{touvron2023llama}.

However, recent studies indicate that unidirectional attention may limit the model's capacity for representation learning. Evidence suggests that bidirectional attention is more effective at integrating contextual information, resulting in improved performance on certain tasks. For example, LLM2Vec \citep{behnamghader2024llm2vec} introduces an additional training phase with a masked token prediction task, preconditioning the model for bidirectional attention. Approaches such as NV-Embed \citep{lee2024nv} and GritLM \citep{muennighoff2024generative} replace unidirectional attention with bidirectional attention during the embedding training phase, often employing mean pooling or more sophisticated latent attention layers to obtain representations for queries and passages.

Despite these advances, we argue that incorporating bidirectional attention during embedding fine-tuning creates a mismatch with the model's pre-training design, potentially undermining its in-context learning and generative properties. To address the trade-off between enhancing embedding representations for specific tasks and preserving the model's inherent generative properties for deep semantic pattern understanding, our approach retains the unidirectional attention mechanism, consistent with the majority of existing embedding methods.

We use the [EOS] token's output embedding as the vector representation for queries and passages, positioning it at the end of inputs to capture both semantic and ICL patterns through causal attention mechanisms, thereby aligning with the foundational pretraining methodology of LLMs.
Specifically, given the tokenized input sequence $\mathrm{T}$: $\mathrm{[BOS]}$, $\mathrm{t}_1$, ..., $\mathrm{t}_N$ is sent into the LLM (Figure \ref{fig:new}):
\begin{equation}
    \mathrm{h}_t = \mathrm{LLM}(\mathrm{T})[\mathrm{EOS}]
\end{equation}
The text embedding is taken from the output embedding of the special token $\mathrm{[EOS]}$.

\subsection{ICL-based Instruction-Tuning}

While previous works \citep{wang2023improving, lee2024nv} have proposed the training method of instruction-tuning, which incorporates a large number of task-specific instructions during the training process, enabling the model to adapt to various downstream retrieval tasks based on different instructions, it is not applicable to the ICL strategy. As demonstrated by GRIT \citep{muennighoff2024generative}, directly supplying few-shot examples when generating embeddings can actually degrade model performance.

To incorporate ICL capabilities into models, we need to modify the conventional instruction tuning strategy. Our approach involves integrating ICL abilities during the training phase. Specifically, we provide task-relevant examples to the query throughout the training process, allowing the model to develop ICL capabilities as it learns.

Recognizing the risk of compromising zero-shot capabilities if examples are consistently provided during training, we propose a dynamic training process. In each training step, queries are supplied with a variable number of few-shot examples, ranging from zero to n, determined by a sampling function. This approach maintains a balance between developing ICL abilities and preserving zero-shot performance.

To further enhance the model's ICL capabilities, we introduce an innovative technique for examples selection. By incorporating in-batch pairs as few-shot examples, we train the model to better differentiate between examples and inputs, aims to improve the model's ability to generate reliable embeddings based on the provided examples.

\section{Experimentens}

In this section, we examine the effectiveness of the ICL training pipeline and reconsider the training methodologies for LLM-based embedding models.

\begin{itemize}
    \item \textbf{RQ 1:} What is the effectiveness of the ICL training strategy for both zero-shot and few-shot learning scenarios?
    \item \textbf{RQ 2:} How does the ICL training strategy impact performance compared to traditional training methods?
    \item \textbf{RQ 3:} How does the integration of in-batch examples affect the performance of the ICL training strategy.
    \item \textbf{RQ 4:} What are the implications of replacing a causal attention mask with a bidirectional attention mask within the framework of LLMs?
    \item \textbf{RQ 5:} What is the impact of various representation strategies, including last token pooling and mean pooling, on model performance?
    \item \textbf{RQ 6:} Do passage-based prompts enhance performance in the ICL training strategy?
\end{itemize}

\subsection{setup}
\label{label:experiments_setup}

\textbf{LLM.}
Following E5-Mistral \citep{wang2023improving}, SFR, and NV-Embedder \citep{lee2024nv}, we have adopted Mistral-7B \citep{jiang2023mistral} as the backbone for our framework.

\textbf{Evaluation.}
We evaluate the performance of our model on MTEB \citep{muennighoff2022mteb} and AIR-Bench. MTEB is a comprehensive benchmark designed to evaluate the performance of text embedding models. AIR-Bench is dedicated to the evaluation of retrieval performance, its testing data is automatically generated by large language models without human intervention.

\textbf{Training Data.}
To ensure a fair comparison, we use the same public datasets from E5-Mistral \citep{wang2023improving}, which includes ELI5 \citep{fan2019eli5}, HotpotQA \citep{yang2018hotpotqa}, FEVER \citep{thorne2018fever}, MIRACL \citep{zhang2023miracl}, MSMARCO passage and document ranking \citep{nguyen2016ms}, NQ \citep{karpukhin2020dense}, NLI \citep{gao2021simcse}, SQuAD \citep{karpukhin2020dense}, TriviaQA \citep{karpukhin2020dense}, Quora Duplicate Questions 
\citep{quora-question-pairs}, MrTyDi \citep{zhang2021mr}, DuReader \citep{qiu2022dureader_retrieval}, and T2Ranking \citep{xie2023t2ranking}, all of which are also used for LLM2Vec \citep{behnamghader2024llm2vec}.

However, methods that typically perform exceptionally well, such as NV-Embedder \citep{lee2024nv} and SFR, often require more training data. Additionally, some of these methods, such as GTE-Qwen2 \citep{li2023towards}, do not disclose their sources of training data. In response, we have developed an enhanced version of our model that leverages a more comprehensive dataset, which includes the following training sets:

\begin{itemize}
    \item \textbf{Retrieval}: ELI5, HotpotQA, FEVER, MSMARCO passage and document ranking, NQ, NLI, SQuAD, TriviaQA, Quora Duplicate Questions, Arguana \citep{wachsmuth2018retrieval}, and FiQA \citep{maia201818}.
    \item \textbf{Reranking}: SciDocsRR \citep{cohan2020specter} and StackOverFlowDupQuestions \citep{liu2018linkso}.
    \item \textbf{Classification}: AmazonReviews-Classification \citep{mcauley2013hidden}, AmazonCounterfactual-Classification \citep{o2021wish}, Banking77-Classification \citep{casanueva2020efficient}, Emotion-Classification \citep{saravia2018carer}, TweetSentimentExtraction-Classification \citep{tweet-sentiment-extraction}, MTOPIntent-Classification \citep{li2020mtop}, IMDB-Classification \citep{maas2011learning}, ToxicConversations-Classification \citep{kaggle2019jigsaw}.
    \item \textbf{Clustering}: \{Arxiv/Biorxiv/Medrxiv/Reddit/StackExchange\}-Clustering-\{S2S/P2P\}, TwentyNewsgroups-Clustering \citep{lang1995newsweeder}.
    \item \textbf{STS}: STS12 \citep{agirre2012semeval}, STS22 \citep{chen2022semeval}, STS-Benchmark \citep{cer2017semeval}.
\end{itemize}

\textbf{Training Detail.}
We fine-tune the Mistral-7B model using a contrastive loss and conduct the process over a single epoch. For efficient fine-tuning, we employ Low-Rank Adaptation (LoRA) \citep{hu2021lora}, setting the LoRA rank to 64 and the LoRA alpha to 32, with a learning rate of 1e-4.
For retrieval tasks, we use in-batch negatives, a strategy not adopted for other tasks. Each dataset incorporates 7 hard negatives. The batch size is set to 512 for retrieval tasks and 256 for other types of tasks. We maintain consistency by using the same dataset throughout one training step, and the maximum sequence length is set at 512 tokens. To distill the score from reranker in retrieval tasks, we use the bge-reranker model as the teacher. For in-context learning training, we implement a randomized sampling method. For each query, we select between 0 to 5 examples from the in-batch training data. The maximum allowable lengths for example queries and documents are set to 256 tokens each, and the combined length for a query with examples is set at 2048 tokens.

\textbf{Evaluation.}
We evaluate the performance of our model under both zero-shot and few-shot conditions. In the few-shot scenario, a consistent set of in-context examples is applied to each query. The examples utilized for evaluation are sourced from training datasets. In cases where training datasets are unavailable, examples are generated using ChatGPT.

\subsection{Main Results}

\begin{table}[!t]
\begin{center}
\begin{small}
\setlength{\tabcolsep}{5pt}
\begin{tabular}{l|ccccccc|c}
\hline
Task & Retr. & Rerank. & Clust. & PairClass. & Class. & STS & Summ. & Avg. \\
\# of datasets $\rightarrow$ & 15 & 4 & 11 & 3 & 12 & 10 & 1 & 56 \\ \hline
\multicolumn{9}{c}{w/ full data} \\ \hline
E5-mistral-7b-instruct & 56.90 & 60.21 & 50.26 & 88.34 & 78.47 & 84.66 & 31.40 & 66.63 \\
GritLM-7B & 57.41 & 60.49 & 50.61 & 87.16 & 79.46 & 83.35 & 30.37 &  66.76  \\
SFR-Embedding & 59.00 & 60.64 & 51.67 & 88.54 & 78.33 & 85.05 & 31.16 & 67.56 \\
Linq-Embed-Mistral & 60.19 & 60.29 & 51.42 & 88.35 & 80.20 & 84.97 & 30.98 & 68.17 \\
voyage-large-2-instruct & 58.28 & 60.09 & 53.35 & 89.24 & 81.49 & 84.31 & 30.84 & 68.23  \\
NV-Embed-v1 & 59.36 & 60.59 & 52.80 & 86.91 & 87.35 & 82.84 & 31.20 & 69.32  \\
bge-multilingual-gemma2 & 59.24 & 59.72 & 54.65 & 85.84 & 88.08 & 83.88 & 31.20 & 69.88 \\
stella\_en\_400M\_v5 & 58.97 & 60.16 & 56.70 & 87.74 & 86.67 & 84.22 & 31.66 & 70.11 \\
gte-Qwen2-7B-instruct & 60.25 & 61.42 & 56.92 & 85.79 & 86.58 & 83.04 & 31.35 & 70.24 \\
SFR-Embedding-2\_R & 60.18 & 60.14 & 56.17 & 88.07 & 89.05 & 81.26 & 30.71 & 70.31 \\
stella\_en\_1.5B\_v5 & 61.01 & 61.21 & 57.69 & 88.07 & 87.63 & 84.51 & 31.49 & 71.19 \\
\textbf{bge-en-icl (zero-shot)} & 61.67 & 59.66 & 57.51 & 86.93 & 88.62 & 83.74 & 30.75 & 71.24 \\
\textbf{bge-en-icl (few-shot)} & 62.16 & 59.82 & 57.89 & 88.14 & 88.95 & 84.24 & 30.77 & \textbf{71.67} \\ \hline
\multicolumn{9}{c}{w/ public data only} \\ \hline
E5-mistral-7b-instruct & 52.78 & 60.38 & 47.78 & 88.47 & 76.80 & 83.77 & 31.90 & 64.56 \\
GritLM-7B & 53.10 & 61.30 & 48.90 & 86.90 & 77.00 & 82.80 & 29.40 & 64.70 \\
% echo-mistral-7b-instruct-lasttoken & 55.52 & 58.14 & 46.32 & 87.34 & 77.43 & 82.56 & 30.73 & 64.68 \\
LLM2Vec-Mistral-supervised & 55.99 & 58.42 & 45.54 & 87.99 & 76.63 & 84.09 & 29.96 & 64.80 \\
\textbf{bge-en-icl (zero-shot)} & 59.59 & 56.85 & 42.61 & 87.87 & 75.47 & 83.30 & 29.52 & 64.67 \\
\textbf{bge-en-icl (few-shot)} & 60.08 & 56.67 & 46.55 & 88.51 & 77.31 & 83.69 & 30.68 & \textbf{66.08} \\
\hline
\end{tabular}
\end{small}
\end{center}
\caption{Top MTEB leaderboard models as of August 27, 2024.}
\label{table:mteb}
% \vspace{-.5cm}
\end{table}

\textbf{MTEB.}
Table \ref{table:mteb} presents the performance of our model, \textit{bge-en-icl}, evaluated on the MTEB benchmark. This evaluation contrasts the results obtained from using the full dataset with those obtained from using only the public dataset. When leveraging the full dataset, our model demonstrates strong capabilities in both zero-shot and few-shot settings, achieving SOTA results in few-shot scenarios.
However, it is important to note that the use of full datasets may introduce inconsistencies, as different models often rely on varying datasets. Notably, many of these models do not disclose the specific datasets they use, leading to potential unfair comparisons.

For a fairer comparison and to better understand the impact of in-context learning, we conducts an evaluation using only the public dataset. Under these constraints, our model's performance in the zero-shot scenario is on par with, or slightly below, that of other models such as LLM2Vec and GritLM. However, in the few-shot settings, our model show significant enhancements (\textcolor{green}{↑1.41}), particularly in the classification and clustering tasks that were not part of the training data. These improvements underscore the potential advantages of in-context learning, emphasizing its efficacy in adapting to tasks beyond the direct scope of initial training parameters. Furthermore, in contrast to training exclusively with public datasets, the utilization of full training data effectively familiarizes the model with these datasets. As a result, the model's ability to generalize effectively is compromised, leading to only a modest improvement in few-shot settings (\textcolor{green}{↑0.43}).

\begin{table}[t!]
\centering
\begin{small}
\setlength{\tabcolsep}{4.5pt}
\begin{tabular}{l|cccccccc|c}
\hline
Domain & wiki & web & news & healthcare & law & finance & arxiv & msmarco & Avg. \\
\# of datasets $\rightarrow$ & 1 & 1 & 1 & 1 & 1 & 1 & 1 & 1 & 8 \\ \hline
\multicolumn{10}{c}{w/ full data} \\ \hline
E5-mistral-7b-instruct & 61.67 & 44.41 & 48.18 & 56.32 & 19.32 & 54.79 & 44.78 & 59.03 & 48.56 \\
SFR-Embedding & 63.46 & 51.27 & 52.21 & 58.76 & 23.27 & 56.94 & 47.75 & 58.99 & 51.58 \\
NV-Embed-v1 & 62.84 & 50.42 & 51.46 & 58.53 & 20.65 & 49.89 & 46.10 & 60.27 & 50.02 \\
Linq-Embed-Mistral & 61.04 & 48.41 & 49.44 & 60.18 & 20.34 & 50.04 & 47.56 & 60.50 & 49.69 \\
gte-Qwen2-7B-instruct & 63.46 & 51.20 & 54.07 & 54.20 & 22.31 & 58.20 & 40.27 & 58.39 & 50.26 \\
stella\_en\_1.5B\_v5 & 61.99 & 50.88 & 53.87 & 58.81 & 23.22 & 57.26 & 44.81 & 61.38 & 51.53 \\
bge-en-icl (zero-shot) & 64.61 & 54.40 & 55.11 & 57.25 & 25.10 & 54.81 & 48.46 & 63.71 & 52.93 \\
bge-en-icl (few-shot) & 64.94 & 55.11 & 56.02 & 58.85 & 28.29 & 57.16 & 50.04 & 64.50 & \textbf{54.36} \\ \hline
\multicolumn{10}{c}{w/ public data only} \\ \hline
bge-en-icl (zero-shot) & 64.82 & 54.96 & 55.82 & 57.06 & 28.87 & 54.46 & 49.60 & 63.25 & 53.60 \\
bge-en-icl (few-shot) & 66.98 & 56.38 & 57.17 & 59.54 & 32.03 & 58.81 & 51.36 & 65.05 & \textbf{55.92} \\ \hline
\end{tabular}
\end{small}
\caption{QA (en, nDCG@10) performance on AIR-Bench 24.04.}
\label{table:air_qa}
\end{table}

\begin{table}[t!]
\centering
\begin{center}
\begin{small}
\setlength{\tabcolsep}{10pt}
\begin{tabular}{l|cccc|c}
\hline
% AIR-Bench\_24.04 & arxiv (4) & book (2) & healthcare (5) & law (4) & Avg. (15) \\ \hline
Domain & arxiv & book & healthcare & law & Avg. \\
\# of datasets $\rightarrow$  & 4 & 2 & 5 & 4 & 15 \\ \hline
\multicolumn{6}{c}{w/ full data} \\ \hline
text-embedding-3-large & 74.53 & 73.16 & 65.83 & 64.47 & 68.77 \\
E5-mistral-7b-instruct & 72.14 & 72.44 & 68.44 & 62.92 & 68.49 \\
SFR-Embedding & 72.79 & 72.41 & 67.94 & 64.83 & 69.00 \\
NV-Embed-v1 & 77.65 & 75.49 & 72.38 & 69.55 & 73.45 \\
Linq-Embed-Mistral & 75.46 & 73.81 & 71.58 & 68.58 & 72.11 \\
gte-Qwen2-7B-instruct & 63.93 & 68.51 & 65.59 & 65.26 & 65.45 \\
stella\_en\_1.5B\_v5 & 73.17 & 74.38 & 70.02 & 69.32 & 71.25 \\
bge-multilingual-gemma2 & 71.77 & 76.46 & 73.96 & 70.86 & 72.88 \\
bge-en-icl (zero-shot) & 78.30 & 78.21 & 73.65 & 67.09 & 73.75 \\
bge-en-icl (few-shot) & 79.63 & 79.36 & 74.80 & 67.79 & \textbf{74.83} \\ \hline
\multicolumn{6}{c}{w/ public data only} \\ \hline
bge-en-icl (zero-shot) & 79.73 & 78.66 & 72.88 & 70.59 & 74.86 \\
bge-en-icl (few-shot) & 79.82 & 80.37 & 74.60 & 71.66 & \textbf{75.98} \\ \hline
\end{tabular}
\end{small}
\end{center}
\caption{Long-Doc (en, Recall@10) performance on AIR-Bench 24.04.}
\label{table:air_longdoc}
\end{table}

\textbf{AIR-Bench.}
The performance of our model is also evaluated using the AIR-Bench dataset. As illustrated in Tables \ref{table:air_qa} and \ref{table:air_longdoc}, the model demonstrates superior performance compared to prior models in both zero-shot and few-shot scenarios, excelling across qa and long-doc tasks. Notably, there is no overlap between the training dataset and the evaluation data for these tasks, highlighting the robustness of the model in scenarios with limited prior exposure. In the few-shot setting, the model exhibits significant improvements over the zero-shot scenario, achieving gains of 1.43 points in the qa task and 1.08 points in the long-doc task. This improvement underscores the efficacy of in-context learning in enhancing the model's generalization capabilities.

However, when the model is trained exclusively using public data, it achieves better results compared to training with the full dataset. This could be attributed to the presence of an excessive amount of MTEB-related data, such as clustering and classification, within the full dataset. Such data might introduce the risk of overfitting, thereby potentially hampering the model's generalization performance on the AIR-Bench dataset.
% Similarly, as all evaluation tasks are out-of-domain, the model trained exclusively on public datasets exhibit significant improvements in few-shot scenarios compared to zero-shot scenarios. Moreover, the model trained on public datasets outperform that trained on full datasets. This discrepancy may be attributed to the inclusion of a wider range of tasks, such as clustering and classification, in the full datasets. Although this wider range of tasks enhances the model's general applicability, it may slightly reduce the model's ability to generalize within specific retrieval domains.
% Further analysis reveals that using only the public dataset does not yield more gains for the few-shot model over the zero-shot scenario, compared to the model using the full dataset, because all tasks are out-of-domain tasks. Nevertheless, there is a noticeable enhancement in overall performance, with scores increasing from 54.36 to 55.92 for the qa task and from 74.83 to 75.98 for the long-doc task.
% % These findings indicate that while specialized tasks like clustering and classification can improve model performance across various domains, they might compromise the model’s overall retrieval abilities.

\subsection{In-context Learning}

\begin{table}[h]
\begin{center}
\begin{small}
\setlength{\tabcolsep}{4.5pt}
\begin{tabular}{l|ccccccc|c}
\hline
% Task & Retr. (15) & Rerank. (4) & Clust. (11) & PairClass. (3) & Class. (12) & STS (10) & Summ. (1) & Avg. (56) \\  \hline
Task & Retr. & Rerank. & Clust. & PairClass. & Class. & STS & Summ. & Avg. \\
\# of datasets $\rightarrow$ & 15 & 4 & 11 & 3 & 12 & 10 & 1 & 56 \\ \hline
\multicolumn{9}{c}{w/ full data} \\ \hline
w/o in-context learning & 59.11 & 57.02 & 42.60 & 87.99 & 76.27 & 83.93 & 30.50 & 64.83 \\
w/ fix examples (zero-shot) & 48.98 & 56.48 & 41.84 & 85.94 & 74.38 & 84.31 & 29.68 & 61.50 \\
w/ fix examples (few-shot) & 59.00 & 56.90 & 45.75 & 88.54 & 75.56 & 84.67 & 30.66 & 65.46 \\
% w/ random examples (zero-shot) &  &  &  &  &  &  &  &  \\
% w/ random examples (few-shot) &  &  &  &  &  &  &  &  \\
w/ in-batch examples (zero-shot) & 59.59 & 56.85 & 42.61 & 87.87 & 75.47 & 83.30 & 29.52 & 64.67 \\
w/ in-batch examples (few-shot) & 60.08 & 56.67 & 46.55 & 88.51 & 77.31 & 83.69 & 30.68 & \textbf{66.08} \\
\hline
\end{tabular}
\end{small}
\end{center}
\caption{Evaluation of various ICL strategies on the MTEB Benchmark.}
\label{table:icl}
\end{table}

To evaluate the impact of the ICL strategy, we conduct a series of ablation studies using the MTEB benchmark. In these studies, we compare the performance of models fine-tuned with the ICL strategy against those fine-tuned without it. Specifically, for ICL training, we employ two distinct training approaches: fixed examples and in-batch examples. In the fixed examples approach, each task was trained using three predetermined examples.
% Conversely, in the random examples approach, the training samples for each query are randomly selected from the entire dataset, rather than from the same batch.

In Table \ref{table:icl}, we present various results from our experiment. When the model is trained without ICL strategy, its average performance is 64.83. This performance is comparable to GritLM \citep{muennighoff2024generative}, LLM2Vec \citep{behnamghader2024llm2vec}, etc. When fixed examples are used during ICL training, there is a significant decline in zero-shot evaluation performance, with a decrease of 3.33 points. This decline is attributed to the model's consistent exposure to specific training examples, which can impair its zero-shot capabilities. On the other hand, in few-shot scenarios, the model demonstrates improved performance, exceeding zero-shot results by 3.96 points and surpassing models trained without ICL by 0.63 points. This also confirms the effectiveness of the ICL strategy.

Meanwhile, the use of in-batch examples, where training may involve zero examples, has preserved the zero-shot capability of the model. There is a modest decrease of 0.16 points compared to the model trained without ICL. Notably, in few-shot scenarios, the performance of the model employing in-batch examples rises to 66.08 (\textcolor{green}{↑1.25}), indicating a robust improvement. Furthermore, when compared with the model utilizing fixed examples, the model trained with in-batch examples displays superior performance in tasks that diverge significantly from the training data, such as classification and clustering tasks.

\subsection{Attention}

\begin{table}[h]
\begin{center}
\begin{small}
\setlength{\tabcolsep}{4.5pt}
\begin{tabular}{l|ccccccc|c}
\hline
% Task & Retr. (15) & Rerank. (4) & Clust. (11) & PairClass. (3) & Class. (12) & STS (10) & Summ. (1) & Avg. (56) \\  \hline
Task & Retr. & Rerank. & Clust. & PairClass. & Class. & STS & Summ. & Avg. \\
\# of datasets $\rightarrow$ & 15 & 4 & 11 & 3 & 12 & 10 & 1 & 56 \\ \hline
\multicolumn{9}{c}{causal attention \& last token pooling} \\ \hline
w/o in-context learning & 59.11 & 57.02 & 42.60 & 87.99 & 76.27 & 83.93 & 30.50 & 64.83 \\
w/ in-context learning (zero-shot) & 59.59 & 56.85 & 42.61 & 87.87 & 75.47 & 83.30 & 29.52 & 64.67 \\
w/ in-context learning (few-shot) & 60.08 & 56.67 & 46.55 & 88.51 & 77.31 & 83.69 & 30.68 & \textbf{66.08} \\ \hline
\multicolumn{9}{c}{causal attention \& mean pooling} \\ \hline
w/o in-context learning & 58.50 & 53.74 & 36.82 & 82.14 & 72.37 & 77.62 & 29.10 & 61.03 \\ \hline
\multicolumn{9}{c}{bidirectional attention \& last token pooling} \\ \hline
w/o in-context learning & 59.59 & 56.96 & 44.34 & 87.61 & 74.77 & 83.81 & 30.12 & 64.96 \\
w/ in-context learning (zero-shot) & 59.77 & 58.09 & 44.04 & 87.87 & 75.35 & 83.97 & 29.75 & 65.19 \\
w/ in-context learning (few-shot) & 60.23 & 57.81 & 44.45 & 88.64 & 77.00 & 83.77 & 29.99 & \textbf{65.74} \\ \hline
\multicolumn{9}{c}{bidirectional attention \& mean pooling} \\ \hline
w/o in-context learning & 59.13 & 57.03 & 43.44 & 87.25 & 75.03 & 84.08 & 29.17 & 64.73 \\
w/ in-context learning (zero-shot) & 59.53 & 57.48 & 43.88 & 88.12 & 74.86 & 83.64 & 29.58 & 64.90 \\
w/ in-context learning (few-shot) & 59.42 & 57.29 & 44.93 & 88.36 & 75.26 & 83.75 & 29.60 & \textbf{65.18} \\
\hline
\end{tabular}
\end{small}
\end{center}
\caption{Results of different attention and pooling mechanisms on the MTEB Benchmark.}
\label{table:bi}
\end{table}

Recent studies have explored modifying causal attention in LLMs to adopt bidirectional attention and employ mean pooling for embedding generation. Notably, models such as GritLM \citep{muennighoff2024generative}, NV-Embed \citep{lee2024nv}, and LLM2Vec \citep{behnamghader2024llm2vec} have utilized these techniques with considerable experimental success. Motivated by these advancements, we explore the potential benefits of implementing bidirectional attention in the ICL scenario. Specifically, we investigate the impacts of various attention and pooling mechanisms, including causal and bidirectional attention, coupled with last token pooling and mean pooling. In a causal attention framework, each token is limited to accessing only preceding tokens' information and not the subsequent ones. Consequently, employing mean pooling tends to yield suboptimal results because of this restriction. We find that the model could not be trained effectively under the ICL setting. Therefore, only results from experiments without ICL are presented in this specific configuration.

Table \ref{table:bi} presents our experimental setup and results in both non-ICL and ICL scenarios. It demonstrates that in non-ICL scenarios, most methods yield consistent performance, aside from the combination of causal attention with mean pooling. In contrast, within ICL scenarios, the integration of causal attention and last token pooling emerges as the superior approach. This configuration appears to resonate with the competencies fostered during the initial training phase of the model, suggesting a strong alignment with the foundational strategies employed during pre-training. Moreover, shifting from causal attention to bidirectional attention does not result in significant improvements, and mean pooling is not necessary for the implementation of bidirectional attention.
% Importantly, transitioning from causal to bidirectional attention might marginally hinder the exploitation of pre-training benefits, thereby achieving positive but not significant few-shot performance. 

Additionally, configurations utilizing bidirectional attention paired with last token pooling are notably effective, excelling in both non-ICL and zero-shot scenarios. This configuration’s performance is also pronounced in few-shot reranking tasks, highlighting its adaptability and potential applicability across various demands.

\subsection{Prompt}

\begin{table}[h]
\begin{center}
\begin{small}
\setlength{\tabcolsep}{4.5pt}
\begin{tabular}{l|ccccccc|c}
\hline
% Task & Retr. (15) & Rerank. (4) & Clust. (11) & PairClass. (3) & Class. (12) & STS (10) & Summ. (1) & Avg. (56) \\  \hline
Task & Retr. & Rerank. & Clust. & PairClass. & Class. & STS & Summ. & Avg. \\
\# of datasets $\rightarrow$ & 15 & 4 & 11 & 3 & 12 & 10 & 1 & 56 \\ \hline
w/o passage prompt (zero-shot) & 59.59 & 56.85 & 42.61 & 87.87 & 75.47 & 83.30 & 29.52 & 64.67 \\
w/o passage prompt (few-shot) & 60.08 & 56.67 & 46.55 & 88.51 & 77.31 & 83.69 & 30.68 & \textbf{66.08} \\
w/ passage prompt (zero-shot) & 59.50 & 46.84 & 39.57 & 81.25 & 71.41 & 80.38 & 30.26 & 61.61 \\
w/ passage prompt (few-shot) & 59.93 & 46.39 & 39.40 & 82.25 & 72.00 & 79.81 & 30.97 & 61.74 \\
\hline
\end{tabular}
\end{small}
\end{center}
\caption{Comparative results of different prompts on the MTEB benchmark.}
\label{table:prompts}
\end{table}

Recently, most LLM-based embedding models have incorporated instruction-based prompts on the query side. However, there has been limited investigation into the efficacy of utilizing prompts on the passage side. To address this gap, our study introduces and explores the use of prompts on the passage side. The specific prompt employed in our study is as follows:
\begin{equation}
    \{\text{passage}\} \newline
    \hspace{0.2cm} \text{Summarize the above passage: }
\end{equation}

Table \ref{table:prompts} presents the results obtained using passage prompts. The results demonstrate that the integration of passage prompts leads to a significant decline in performance across all tasks except retrieval. This indicates that further exploration and experimentation are needed when employing prompts at the passage level.

% - distill train
% - distill + bi + mean train
% - distill + latent + bi train
% - distill + ICL train
%             --fix examples (3 examples) for each task
%             --in-batch examples
% - distill + bi + ICL + mean train

% \subsection{directional}

% \subsection{pooling}

% \subsection{instrcution}

\section{Conclusion}

% In this paper, we rethink the application of LLMs in dense retrieval and investigate potential avenues for their optimization when utilized as embedding models.
In this paper, we explore the utilization of in-context learning (ICL) derived from large language models (LLMs) for generating text embeddings and investigate various methods of LLMs as embedding models. Specifically, we examine the integration of attention mechanisms, different pooling methods, and passage prompts. We advocate for maintaining the model's original architecture while embedding in-context learning capabilities into the dense retrieval process. Our approach necessitates no modifications to the model's architecture; instead, it involves altering the prompt on the query side to include in-context learning features in the embedding generation task. Despite its simplicity, our method proves highly effective on the MTEB and AIR-Bench benchmarks.

\bibliography{iclr2025_conference}
\bibliographystyle{iclr2025_conference}

\newpage
\appendix
\section{Instruction}

\begin{table}[h]
\begin{center}
\begin{small}
\setlength{\tabcolsep}{4.5pt}
\setlength{\extrarowheight}{2pt}
\resizebox{\textwidth}{!}{
\begin{tabular}{ll}
\toprule
Task Name & Instruction Template \\ \midrule
ArguAna    &     Given a claim, find documents that refute the claim. \\
ClimateFEVER    &     \begin{tabular}[c]{@{}l@{}}Given a claim about climate change, retrieve documents that support or refute \\ the claim.\end{tabular} \\
CQADupStack    &   \begin{tabular}[c]{@{}l@{}}Given a question, retrieve detailed question descriptions from Stackexchange that are \\ duplicates to the given question.\end{tabular} \\
DBPedia    &     Given a query, retrieve relevant entity descriptions from DBPedia. \\
FEVER    &     Given a claim, retrieve documents that support or refute the claim. \\
FiQA2018    &     Given a financial question, retrieve user replies that best answer the question. \\
HotpotQA    &     Given a multi-hop question, retrieve documents that can help answer the question. \\
MSMARCO    &     Given a web search query, retrieve relevant passages that answer the query. \\
NFCorpus    &     Given a question, retrieve relevant documents that best answer the question. \\
Natural Question    &     Given a question, retrieve Wikipedia passages that answer the question. \\
QuoraRetrieval    &     \begin{tabular}[c]{@{}l@{}}Given a question, retrieve questions that are semantically equivalent to the given \\ question.\end{tabular} \\
SCIDOCS    &     Given a scientific paper title, retrieve paper abstracts that are cited by the given paper. \\
SciFact    &     Given a scientific claim, retrieve documents that support or refute the claim. \\
Touche2020    &     Given a question, retrieve detailed and persuasive arguments that answer the question. \\
TREC-COVID    &     Given a query, retrieve documents that answer the query. \\ 
STS*    &     Retrieve semantically similar text. \\
SummEval    &     Given a news summary, retrieve other semantically similar summaries. \\
AmazonCounterfactualClassification    &     \begin{tabular}[c]{@{}l@{}}Classify a given Amazon customer review text as either counterfactual \\ or not-counterfactual.\end{tabular} \\
AmazonPolarityClassification    &     Classify Amazon reviews into positive or negative sentiment. \\
AmazonReviewsClassification    &     Classify the given Amazon review into its appropriate rating category. \\
Banking77Classification    &     Given a online banking query, find the corresponding intents. \\
EmotionClassification    &   \begin{tabular}[c]{@{}l@{}}Classify the emotion expressed in the given Twitter message into one of the six \\ emotions: anger, fear, joy, love, sadness, and surprise.\end{tabular}   \\
    % &     \qquad into one of the six emotions: anger, fear, joy, love, sadness, and surprise. \\  
ImdbClassification    &     \begin{tabular}[c]{@{}l@{}}Classify the sentiment expressed in the given movie review text from \\ the IMDB dataset.\end{tabular} \\
MassiveIntentClassification    &     Given a user utterance as query, find the user intents. \\
MassiveScenarioClassification    &     Given a user utterance as query, find the user scenarios. \\
MTOPDomainClassification    &     Classify the intent domain of the given utterance in task-oriented conversation. \\
MTOPIntentClassification    &     Classify the intent of the given utterance in task-oriented conversation. \\
ToxicConversationsClassification    &     Classify the given comments as either toxic or not toxic. \\
TweetSentimentExtractionClassification    &     Classify the sentiment of a given tweet as either positive, negative, or neutral. \\
ArxivClusteringP2P    &     \begin{tabular}[c]{@{}l@{}}Identify the main and secondary category of Arxiv papers based on the titles \\ and abstracts.\end{tabular} \\
ArxivClusteringS2S    &     Identify the main and secondary category of Arxiv papers based on the titles. \\
BiorxivClusteringP2P    &     Identify the main category of Biorxiv papers based on the titles and abstracts. \\
BiorxivClusteringS2S    &     Identify the main category of Biorxiv papers based on the titles. \\
MedrxivClusteringP2P    &     Identify the main category of Medrxiv papers based on the titles and abstracts. \\
MedrxivClusteringS2S    &     Identify the main category of Medrxiv papers based on the titles. \\
RedditClustering    &     Identify the topic or theme of Reddit posts based on the titles. \\
RedditClusteringP2P    &     Identify the topic or theme of Reddit posts based on the titles and posts. \\
StackExchangeClustering    &     Identify the topic or theme of StackExchange posts based on the titles. \\
StackExchangeClusteringP2P    &     Identify the topic or theme of StackExchange posts based on the given paragraphs. \\
TwentyNewsgroupsClustering    &     Identify the topic or theme of the given news articles. \\
AskUbuntuDupQuestions    &     Retrieve duplicate questions from AskUbuntu forum. \\
MindSmallReranking    &     Retrieve relevant news articles based on user browsing history. \\
SciDocsRR    &     Given a title of a scientific paper, retrieve the titles of other relevant papers. \\
StackOverflowDupQuestions    &     Retrieve duplicate questions from StackOverflow forum. \\
SprintDuplicateQuestions    &     Retrieve duplicate questions from Sprint forum. \\
TwitterSemEval2015    &     Retrieve tweets that are semantically similar to the given tweet. \\
TwitterURLCorpus    &     Retrieve tweets that are semantically similar to the given tweet. \\ \midrule
AIR-Bench    &     Given a question, retrieve passages that answer the question. \\ \bottomrule
\end{tabular}
}
\end{small}
\end{center}
\caption{The instruction we used on the MTEB and AIR-Bench benchmarks.}
% , where all STS tasks use the same instruction as described in ``STS*''. These instructions are adopted from E5-Mistral-7B \citep{wang2023improving}. To ensure sentence completeness, we add a period at the end.}
\label{table:evalinstructions}
\end{table}

\newpage

\section{Detailed MTEB Results}

\begin{table}[h]
\begin{center}
\hspace*{-1.6cm}
\begin{small}\begin{tabular}{l|ccccc|cccc}
\hline
Dataset & \begin{tabular}[c]{@{}l@{}}NV-Em\\ bed-v1\end{tabular} & \begin{tabular}[c]{@{}l@{}}bge-multilin\\ gual-gemma2\end{tabular} & \begin{tabular}[c]{@{}l@{}}gte-Qwen2-\\ 7B-instruct\end{tabular} & \begin{tabular}[c]{@{}l@{}}SFR-Embe\\ dding-2\_R\end{tabular} & \begin{tabular}[c]{@{}l@{}}stella\_en\_\\ 1.5B\_v5\end{tabular} & \begin{tabular}[c]{@{}l@{}}bge-en-icl\\ (zero-shot)\end{tabular} & \begin{tabular}[c]{@{}l@{}}bge-en-icl\\ (few-shot)\end{tabular} \\ \hline
ArguAna & 68.21 & 77.37 & 64.27 & 62.34 & 65.27 & 82.76 & 83.08 \\
ClimateFEVER & 34.72 & 39.37 & 45.88 & 34.43 & 46.11 & 45.35 & 45.43 \\
CQADupStack & 50.51 & 47.94 & 46.43 & 46.11 & 47.75 & 47.23 & 47.31 \\
DBPEDIA & 48.29 & 51.37 & 52.42 & 51.21 & 52.28 & 50.42 & 51.63 \\
FEVER & 87.77 & 90.38 & 95.11 & 92.16 & 94.83 & 91.96 & 92.83 \\
FiQA2018 & 63.10 & 60.04 & 62.03 & 61.77 & 60.48 & 58.77 & 59.67 \\
HotpotQA & 79.92 & 83.26 & 73.08 & 81.36 & 76.67 & 84.98 & 85.14 \\
MSMARCO & 46.49 & 45.71 & 45.98 & 42.18 & 45.22 & 46.72 & 46.79 \\
NFCorpus & 38.04 & 38.11 & 40.60 & 41.34 & 42.00 & 40.69 & 41.85 \\
Natural Question & 71.22 & 71.45 & 67.00 & 73.96 & 71.80 & 73.85 & 73.88 \\
QuoraRetrieval & 89.21 & 90.04 & 90.09 & 89.58 & 90.03 & 91.02 & 90.95 \\
SCIDOCS & 20.19 & 26.93 & 28.91 & 24.87 & 26.64 & 25.25 & 25.26 \\
SciFact & 78.43 & 72.05 & 79.06 & 85.91 & 80.09 & 78.33 & 79.09 \\
Touche2020 & 28.38 & 30.26 & 30.57 & 28.18 & 29.94 & 29.67 & 30.48 \\
TREC-COVID & 85.88 & 64.27 & 82.26 & 87.28 & 85.98 & 78.11 & 79.08 \\
BIOSSES & 85.59 & 85.74 & 81.37 & 87.60 & 83.11 & 86.35 & 86.47 \\
SICK-R & 82.80 & 82.66 & 79.28 & 77.01 & 82.89 & 83.87 & 83.87 \\
STS12 & 76.22 & 77.71 & 79.55 & 75.67 & 80.09 & 77.73 & 78.14 \\
STS13 & 86.30 & 87.45 & 88.83 & 82.40 & 89.68 & 85.98 & 86.59 \\
STS14 & 82.09 & 83.48 & 83.87 & 79.93 & 85.07 & 82.34 & 82.83 \\
STS15 & 87.24 & 87.63 & 88.54 & 85.82 & 89.39 & 87.35 & 87.77 \\
STS16 & 84.77 & 86.70 & 86.49 & 84.50 & 87.15 & 86.54 & 87.04 \\
STS17 & 87.42 & 91.18 & 88.73 & 88.93 & 91.35 & 91.25 & 91.25 \\
STS22 & 69.85 & 69.02 & 66.88 & 67.10 & 68.10 & 68.08 & 70.07 \\
STSBenchmark & 86.14 & 87.25 & 86.85 & 83.60 & 88.23 & 87.92 & 88.42 \\
SummEval & 31.20 & 31.20 & 31.35 & 30.71 & 31.49 & 30.75 & 30.77 \\
SprintDuplicateQuestions & 95.94 & 90.94 & 92.82 & 97.62 & 96.04 & 95.06 & 97.23 \\
TwitterSemEval2015 & 78.73 & 79.64 & 77.96 & 78.57 & 80.58 & 78.54 & 79.34 \\
TwitterURLCorpus & 86.05 & 86.95 & 86.59 & 88.03 & 87.58 & 87.19 & 87.84 \\
AmazonCounterfactual & 95.12 & 89.48 & 91.31 & 92.72 & 92.87 & 92.88 & 93.15 \\
AmazonPolarity & 97.14 & 96.90 & 97.50 & 97.31 & 97.16 & 96.86 & 96.98 \\
AmazonReviews & 55.47 & 61.60 & 62.56 & 61.04 & 59.36 & 61.28 & 61.46 \\
Banking77 & 90.34 & 92.53 & 87.57 & 90.02 & 89.79 & 91.42 & 91.49 \\
Emotion & 91.71 & 92.97 & 79.45 & 93.37 & 84.29 & 93.31 & 93.36 \\
Imdb & 97.06 & 96.66 & 96.75 & 96.80 & 96.66 & 96.91 & 96.91 \\
MassiveIntent & 80.07 & 82.05 & 85.41 & 85.97 & 85.83 & 82.26 & 82.93 \\
MassiveScenario & 81.74 & 84.40 & 89.77 & 90.61 & 90.20 & 83.92 & 85.60 \\
MTOPDomain & 96.51 & 98.61 & 99.04 & 98.58 & 99.01 & 97.99 & 98.42 \\
MTOPIntent & 89.77 & 95.51 & 91.88 & 91.30 & 92.78 & 93.56 & 94.00 \\
ToxicConversations & 92.60 & 87.34 & 85.12 & 91.14 & 88.76 & 93.16 & 93.17 \\
TweetSentimentExtraction & 80.60 & 78.86 & 72.58 & 79.70 & 74.84 & 79.90 & 79.93 \\
Arxiv-P2P & 53.76 & 54.91 & 54.46 & 54.02 & 55.44 & 54.42 & 54.44 \\
Arxiv-S2S & 49.59 & 50.28 & 51.74 & 48.82 & 50.66 & 49.17 & 49.33 \\
Biorxiv-P2P & 48.15 & 52.64 & 50.09 & 50.76 & 50.68 & 52.32 & 53.05 \\
Biorxiv-S2S & 44.74 & 49.20 & 46.65 & 46.57 & 46.87 & 48.38 & 48.38 \\
Medrxiv-P2P & 39.24 & 45.81 & 46.23 & 46.66 & 46.87 & 46.13 & 45.86 \\
Medrxiv-S2S & 36.98 & 44.11 & 44.13 & 44.18 & 44.65 & 44.20 &  44.33\\
Reddit & 63.20 & 56.03 & 73.55 & 62.92 & 72.86 & 71.20 & 72.33 \\
Reddit-P2P & 68.01 & 65.83 & 74.13 & 72.74 & 75.27 & 72.17 & 72.72 \\
StackExchange & 74.99 & 66.21 & 79.86 & 76.48 & 80.29 & 81.29 & 81.32 \\
StackExchange-P2P & 42.04 & 45.74 & 49.41 & 48.29 & 49.57 & 45.53 & 46.05 \\
TwentyNewsgroups & 60.13 & 70.44 & 53.91 & 66.42 & 61.43 & 68.51 & 68.98 \\
AskUbuntuDupQuestions & 67.50 & 64.59 & 67.58 & 66.71 & 67.33 & 64.80 & 65.15 \\
MindSmallRerank & 30.82 & 31.79 & 33.36 & 31.26 & 33.05 & 30.60 & 30.60 \\
SciDocsRR & 87.26 & 87.60 & 89.09 & 87.29 & 89.20 & 86.90 & 86.96 \\
StackOverflowDupQuestions & 56.58 & 54.90 & 55.66 & 55.32 & 55.25 & 56.32 & 56.71 \\ \hline
\bf{MTEB Average (56)} & 69.32 & 69.88 & 70.24 & 70.31 & 71.19 & 71.24 & \textbf{71.67} \\ \hline
\end{tabular}
\end{small}
\end{center}
\vspace{-0.4cm}
\caption{MTEB results with full data.}
\label{table:fullmteb}
\end{table}

\begin{table}[h]
\begin{center}
% \hspace*{-1.6cm}
\begin{small}\begin{tabular}{l|cc}
\hline
Dataset & bge-en-icl (zero-shot) & bge-en-icl (few-shot) \\ \hline
ArguAna & 55.81 & 55.41 \\
ClimateFEVER & 45.17 & 45.14 \\
CQADupStack & 46.03 & 46.46 \\
DBPEDIA & 50.79 & 51.14 \\
FEVER & 91.96 & 92.42 \\
FiQA2018 & 58.49 & 58.15 \\
HotpotQA & 84.34 & 84.68 \\
MSMARCO & 46.52 & 46.56 \\
NFCorpus & 40.16 & 40.96 \\
Natural Question & 73.56 & 74.01 \\
QuoraRetrieval & 90.79 & 90.89 \\
SCIDOCS & 20.56 & 20.87 \\
SciFact & 78.10 & 79.65 \\
Touche2020 & 33.64 & 34.93 \\
TREC-COVID & 77.89 & 79.95 \\
BIOSSES & 86.80 & 87.49 \\
SICK-R & 83.83 & 83.69 \\
STS12 & 77.80 & 78.39 \\
STS13 & 84.90 & 85.62 \\
STS14 & 82.53 & 82.62 \\
STS15 & 88.33 & 88.52 \\
STS16 & 86.14 & 86.44 \\
STS17 & 91.65 & 91.79 \\
STS22 & 63.79 & 64.83 \\
STSBenchmark & 87.27 & 87.52 \\
SummEval & 29.52 & 30.68 \\
SprintDuplicateQuestions & 94.79 & 96.09 \\
TwitterSemEval2015 & 81.53 & 82.04 \\
TwitterURLCorpus & 87.30 & 87.39 \\
AmazonCounterfactual & 80.78 & 83.36 \\
AmazonPolarity & 88.57 & 92.69 \\
AmazonReviews & 47.55 & 49.85 \\
Banking77 & 87.57 & 88.70 \\
Emotion & 54.29 & 54.24 \\
Imdb & 81.14 & 84.96 \\
MassiveIntent & 78.54 & 79.24 \\
MassiveScenario & 79.27 & 82.00 \\
MTOPDomain & 95.57 & 96.61 \\
MTOPIntent & 85.32 & 88.19 \\
ToxicConversations & 63.58 & 64.68 \\
TweetSentimentExtraction & 63.47 & 63.16 \\
Arxiv-P2P & 47.22 & 48.97 \\
Arxiv-S2S & 42.87 & 45.35 \\
Biorxiv-P2P & 33.17 & 38.37 \\
Biorxiv-S2S & 35.00 & 37.05 \\
Medrxiv-P2P & 28.74 & 30.24 \\
Medrxiv-S2S & 28.10 & 31.45 \\
Reddit & 53.83 & 59.14 \\
Reddit-P2P & 64.40 & 65.51 \\
StackExchange & 57.50 & 68.61 \\
StackExchange-P2P & 34.21 & 36.01 \\
TwentyNewsgroups & 43.65 & 51.40 \\
AskUbuntuDupQuestions & 63.71 & 62.96 \\
MindSmallRerank & 27.90 & 27.90 \\
SciDocsRR & 84.31 & 84.24 \\
StackOverflowDupQuestions & 51.48 & 51.56 \\ \hline
\bf{MTEB Average (56)} & 64.67 & 66.08 \\ \hline
\end{tabular}
\end{small}
\end{center}
\vspace{-0.4cm}
\caption{MTEB results with public data only.}
\label{table:fullmteb_2}
\end{table}

\clearpage
\newpage
\section{Multilingual Embedding Model}
\label{multilingual}

Considering that the LLM-based multilingual embedding models are still relatively scare, we further train a LLM-based multilingual embedding model, \textit{bge-multilingual-gemma2}, on a diverse range of languages and tasks. It is noted that bge-multilingual-gemma2 is just our initial attempt, and we have not explored the in-context learning (ICL) capabilities of bge-multilingual-gemma2. The exploration of ICL capabilities in the multilingual embedding models is probably a future research topic. However, in our experiment, the new multilingual embedding model has already achieved excellent performance on multiple embedding benchmarks, and especially led to new state-of-the-art results on several multilingual benchmarks.

\subsection{Setup}

\textbf{LLM}. XLM-RoBERTa~\citep{conneau-etal-2020-unsupervised} demonstrated that the larger vocabulary size were beneficial for improving the multilingual capability of language models. Therefore, we employ Gemma-2-9b~\citep{team2024gemma} as the backbone for the new multilingual embedding model, considering that the vocabulary size of Gemma-2-9b is 256K, which is larger than the vocabulary size of other LLMs, such as Qwen2~\citep{qwen2} or Llama 3~\citep{dubey2024llama3herdmodels}.

\textbf{Dataset}. In addition to MTEB~\citep{muennighoff2022mteb} and AIR-Bench~\footnote{\href{https://github.com/AIR-Bench/AIR-Bench}{https://github.com/AIR-Bench/AIR-Bench}}, we also evaluate the multilingual capability of bge-multilingual-gemma2 on MIRACL~\citep{zhang2023miracl}, FR-MTEB~\citep{ciancone2024mtebfrenchresourcesfrenchsentence}, PL-MTEB~\citep{poswiata2024plmteb} and C-MTEB~\citep{xiao2024c}.

\textbf{Training Data}. For the Engilsh training data, we use most of the datasets used by bge-en-icl. For the Chinese training data, in addition to the datasets used by BGE-M3~\citep{chen2024bge}, more datasets in retrieval, classification, and clustering tasks are included. For the multilingual training data, we still use the two multilingual datasets used by BGE-M3. The full training data used by bge-multilingual-gemma2 includes:

\begin{itemize}
    \item \textbf{English}: The English datasets (refer to Section~\ref{label:experiments_setup}) used by bge-en-icl are employed, except for the MSMARCO document ranking dataset.
    \item \textbf{Chinese}: The Chinese datasets used by BGE-M3~\citep{chen2024bge} are employed. The retrieval datasets including the three domain-specific datasets in Multi-CPR~\citep{multi_cpr}, the classification datasets including AmazonReviews-Classification \citep{mcauley2013hidden} and MultilingualSentiment-Classification~\citep{mollanorozy-etal-2023-cross}, and the clustering datasets including CSL-Clustering-\{S2S/P2P2\}~\citep{li2022csl} are addtionally employed.
    \item \textbf{Multilingual}: Two multilingual retrieval datasets including MIRACL \citep{zhang2023miracl} and Mr.TyDi \citep{zhang2021mr} are employed.
\end{itemize}

\textbf{Training Detail}. We fine-tune the Gemma-2-9b model using a contrastive loss and conduct the process over a single epoch. For efficient fine-tuning, we employ Low-Rank Adaptation (LoRA) \citep{hu2021lora}, setting the LoRA rank to 64 and the LoRA alpha to 32, with a learning rate of 1e-4. We use in-batch negatives only for retrieval tasks, where each dataset incorporates 7 hard negatives. For the retrieval tasks and the other tasks, we set the batch size to 512 and 256, respectively. We consistently use the same dataset throughout one step, and the maximum sequence length remains capped at 512 tokens. Meanwhile, we use bge-reranker as the teacher to distill our model in retrieval tasks.

\textbf{Evaluation}. On the MTEB benchmark, the instructions used by bge-multilingual-gemma2 are consistent with the instructions used by bge-en-icl, which are shown in Table~\ref{table:evalinstructions}. The instructions used by bge-multilingual-gemma2 on the C-MTEB, PL-MTEB and FR-MTEB benchmarks are available in Table~\ref{table:multilingual_evalinstructions}. On the MIRACL benchmark, we use the same instruction for all 18 languages: ``Given a question, retrieve Wikipedia passages that answer the question.''. On the AIR-Bench benchmark, for the sake of simplicity, we also use the same instruction for all datasets: ``Given a question, retrieve passages that answer the question.''.

\subsection{Main Results}

\textbf{MIRACL}. Following BGE-M3~\citep{chen2024bge}, we evaluate the multilingual retrieval performance with MIRACL~\citep{zhang2023miracl}. We cite most of the results reported in the Table~1 of BGE-M3's paper. It should be noted that the results of BM25 are lower than the results reported in MIRACL's paper, as the BM25 tested in BGE-M3's paper used the same tokenizer with BGE-M3. We also include another recent work mGTE~\citep{zhang2024mgtegeneralizedlongcontexttext} as one of the baseline models.

As shown in Table~\ref{table:miracl_ndcg_results}, our model \textit{bge-multilingual-gemma2} achieves the state-of-the-art (SOTA) performance in all 18 languages. The overall performance of bge-multilingual-gemma2 is 74.1, which is far ahead of the 
performance of the previous best model BGE-M3 (Dense), indicating the excellent multilingual retrieval capability of bge-multilingual-gemma2. The results of Recall@100 are available in Table~\ref{table:miracl_recall_results}.

\begin{table*}[!t]
    \centering
    \setlength{\tabcolsep}{1.5pt}
    \setlength{\extrarowheight}{3pt}
    \resizebox{\textwidth}{!}{
        \begin{tabular}{l|c|cccccccccccccccccc}
        % \ChangeRT{1pt} 
        \hline
        Model & Avg. & ar & bn & en & es & fa & fi & fr & hi & id & ja & ko & ru & sw & te & th & zh & de & yo \\
        \hline
        % \multicolumn{20}{l}{Baselines~(\textit{Prior Work})} \\
        % \hline
        % BM25 & 38.5 & 48.1 & 50.8 & 35.1 & 31.9 & 33.3 & 55.1 & 18.3 & 45.8 & 44.9 & 36.9 & 41.9 & 33.4 & 38.3 & 49.4 & 48.4 & 18.0 & 22.6 & 40.6 \\
        BM25 & 31.9 & 39.5 & 48.2 & 26.7 & 7.7 & 28.7 & 45.8 & 11.5 & 35.0 & 29.7 & 31.2 & 37.1 & 25.6 & 35.1 & 38.3 & 49.1 & 17.5 & 12.0 & 56.1 \\
        mDPR & 41.8 & 49.9 & 44.3 & 39.4 & 47.8 & 48.0 & 47.2 & 43.5 & 38.3 & 27.2 & 43.9 & 41.9 & 40.7 & 29.9 & 35.6 & 35.8 & 51.2 & 49.0 & 39.6  \\
        mContriever & 43.1 & 52.5 & 50.1 & 36.4 & 41.8 & 21.5 & 60.2 & 31.4 & 28.6 & 39.2 & 42.4 & 48.3 & 39.1 & 56.0 & 52.8 & 51.7 & 41.0 & 40.8 & 41.5 \\
        % mE5$_{\mathrm{{\text{large}}}}$ 
        mE5$_{\mathrm{{\text{large}}}}$ & 66.6 & 76.0 & 75.9 & 52.9 & 52.9 & 59.0 & 77.8 & 54.5 & 62.0 & 52.9 & 70.6 & 66.5 & 67.4 & 74.9 & 84.6 & 80.2 & 56.0 & 56.4 & 78.3 \\
        % E5$_{\mathrm{\text{mistral-7b}}}$ 
        E5$_{\mathrm{\text{mistral-7b}}}$ & 63.4 & 73.3 & 70.3 & 57.3 & 52.2 & 52.1 & 74.7 & 55.2 & 52.1 & 52.7 & 66.8 & 61.8 & 67.7 & 68.4 & 73.9 & 74.0 & 54.0 & 54.1 & 79.7 \\
        OpenAI-3 & 54.9 & - & - & - & - & - & - & - & - & - & - & - & - & - & - & - & - & - & - \\
        BGE-M3 (Dense) & 69.2 & 78.4 & 80.0 & 56.9 & 56.1 & 60.9 & 78.6 & 58.3 & 59.5 & 56.1 & 72.8 & \textbf{69.9} & 70.1 & 78.7 & 86.2 & 82.6 & 62.7 & 56.7 & 81.8 \\
        mGTE-TRM (Dense) & 62.1 & 71.4 & 72.7 & 54.1 & 51.4 & 51.2 & 73.5 & 53.9 & 51.6 & 50.3 & 65.8 & 62.7 & 63.2 & 69.9 & 83.0 & 74.0 & 60.8 & 49.7 & 58.3 \\
        % \hline
        % \multicolumn{20}{l}{bge-multilingual-gemma2~(\textit{Our Work})} \\
        % \hline
        \textbf{bge-multilingual-gemma2} & \textbf{74.1} & \textbf{81.0} & \textbf{82.3} & \textbf{64.5} & \textbf{64.2} & \textbf{64.0} & \textbf{81.2} & \textbf{64.2} & \textbf{68.2} & \textbf{61.5} & \textbf{79.1} & 69.7 & \textbf{77.0} & \textbf{81.9} & \textbf{88.1} & \textbf{84.6} & \textbf{68.0} & \textbf{63.5} & \textbf{90.3} \\
        \hline
        \end{tabular}
    }
    \vspace{-5pt}
    \caption{Multi-lingual retrieval performance on the MIRACL~\citep{zhang2023miracl} dev set (measured by nDCG@10).}
    \label{table:miracl_ndcg_results}
\end{table*}

\begin{table*}[!t]
    \centering
    \setlength{\tabcolsep}{1.5pt}
    \setlength{\extrarowheight}{3pt}
    \resizebox{\textwidth}{!}{
        \begin{tabular}{l|c|cccccccccccccccccc}
        \hline
        Model & Avg. & ar & bn & en & es & fa & fi & fr & hi & id & ja & ko & ru & sw & te & th & zh & de & yo \\
        \hline
        % BM25 (D.T.) & 77.2 & 88.9 & 90.9 & 81.9 & 70.2 & 73.1 & 89.1 & 65.3 & 86.8 & 90.4 & 80.5 & 78.3 & 66.1 & 70.1 & 83.1 & 88.7 & 56.0 & 57.2 & 73.3 \\
        BM25 & 67.3 & 78.7 & 90.0 & 63.6 & 25.4 & 68.1 & 81.2 & 50.2 & 73.8 & 71.8 & 73.6 & 70.1 & 56.4 & 69.9 & 73.3 & 87.5 & 55.1 & 42.8 & 80.1 \\
        mDPR & 79.0 & 84.1 & 81.9 & 76.8 & 86.4 & 89.8 & 78.8 & 91.5 & 77.6 & 57.3 & 82.5 & 73.7 & 79.7 & 61.6 & 76.2 & 67.8 & 94.4 & 89.8 & 71.5  \\
        mContriever & 84.9 & 92.5 & 92.1 & 79.7 & 84.1 & 65.4 & 95.3 & 82.4 & 64.6 & 80.2 & 87.8 & 87.5 & 85.0 & 91.1 & 96.1 & 93.6 & 90.3 & 84.1 & 77.0 \\
        mE5$_{\mathrm{{\text{large}}}}$ & 94.1 & 97.3 & 98.2 & 87.6 & 89.1 & 92.9 & 98.1 & 90.6 & 93.9 & 87.9 & 97.1 & 93.4 & 95.5 & 96.7 & 99.2 & 98.9 & 93.3 & 90.7 & 93.1 \\
        E5$_{\mathrm{\text{mistral-7b}}}$ & 92.7 & 96.0 & 96.0 & 90.2 & 87.5 & 88.0 & 96.7 & 92.8 & 89.9 & 88.4 & 95.1 & 89.4 & 95.0 & 95.5 & 95.1 & 96.5 & 90.1 & 88.7 & 97.9 \\
        % OpenAI-embebedding-3 & 54.9 \\
        BGE-M3 (Dense) & 95.5 & 97.6 & 98.7 & 90.7 & 91.1 & 94.0 & 97.9 & 93.8 & 94.4 & 90.5 & 97.5 & \textbf{95.5} & 95.9 & 97.2 & 99.4 & 99.1 & 96.9 & 90.9 & 98.7 \\
        bge-multilingual-gemma2 & \textbf{97.2} & \textbf{99.0} & \textbf{98.9} & \textbf{95.4} & \textbf{94.5} & \textbf{95.0} & \textbf{98.5} & \textbf{96.2} & \textbf{96.5} & \textbf{95.3} & \textbf{98.9} & 95.4 & \textbf{98.0} & \textbf{98.0} & \textbf{99.7} & \textbf{99.6} & \textbf{97.2} & \textbf{94.1} & \textbf{100.0} \\
        \hline
        \end{tabular}
    }
    \vspace{-5pt}
    \caption{Multi-lingual retrieval performance on the MIRACL~\citep{zhang2023miracl} dev set (measured by Recall@100).}
    \vspace{-10pt}
    \label{table:miracl_recall_results}
\end{table*}

\textbf{FR-MTEB} \& \textbf{PL-MTEB} \& \textbf{C-MTEB}. We further evaluate our model on FR-MTEB~\citep{ciancone2024mtebfrenchresourcesfrenchsentence}, PL-MTEB~\citep{poswiata2024plmteb} and C-MTEB~\citep{xiao2024c} benchmarks. FR-MTEB consists of 26 datasets in 6 different tasks, PL-MTEB consists of 26 datasets in 5 different tasks, and C-MTEB consists of 35 datasets in 6 different tasks. We use the API provided by MTEB~\footnote{\href{https://github.com/embeddings-benchmark/mteb}{https://github.com/embeddings-benchmark/mteb}} to perform evaluation.

The results shown in Table~\ref{table:fr_mteb_results}, Table~\ref{table:pl_mteb_results} and Table~\ref{table:c_mteb_results} are all available in the MTEB leaderboard~\footnote{\href{https://huggingface.co/spaces/mteb/leaderboard}{https://huggingface.co/spaces/mteb/leaderboard}}. We can observe that bge-multilingual-gemma2 leads to new SOTA performances on both FR-MTEB and PL-MTEB benchmarks, and especially achieves very excellent results in the retrieval tasks (Retr.). On the C-MTEB benchmark, bge-multilingual-gemma2 surpasses most of the baseline models, such as e5-mistral-7b-instruct, bge-large-zh-v1.5, etc. However, its overall performance on C-MTEB benchmark is still slightly worse than gte-Qwen2-7B-instruct, which can be attributed to Gemma-2's Chinese proficiency being worse than that of Qwen2.

\begin{table}[!t]
    \centering
    \resizebox{\textwidth}{!}{
        \begin{tabular}{l|ccccccc|c}
        \hline
        Task & Retr. & Rerank. & Clust. & PairClass. & Class. & STS & Summ. & Avg. \\  
        \# of datasets $\rightarrow$ & 5 & 2 & 7 & 2 & 6 & 3 & 1 & 26 \\ \hline
        mistral-embed & 46.81 & 80.46 & 44.74 & 77.32 & 68.61 & 79.56 & \textbf{31.47} & 59.41 \\
        gte-multilingual-base & 52.97 & 76.47 & 41.66 & 79.46 & 68.72 & 81.36 & 29.74 & 59.79 \\
        voyage-multilingual-2 & 54.56 & 82.59 & 46.57 & 78.66 & 68.56 & 80.13 & 29.96 & 61.65 \\
        gte-Qwen2-1.5B-instruct & 52.56 & 83.76 & 55.01 & 86.88 & 78.02 & 81.26 & 30.5 & 66.6 \\
        gte-Qwen2-7B-instruct & 55.65 & 78.7 & 55.56 & \textbf{90.43} & \textbf{81.76} & 82.31 & 31.45 & 68.25 \\
        \textbf{bge-multilingual-gemma2} & \textbf{63.47} & \textbf{85.22} & \textbf{56.48} & 85.07 & 81.62 & \textbf{82.59} & 31.26 & \textbf{70.08} \\
        \hline
        \end{tabular}
    }
\vspace{-5pt}
\caption{Results on the FR-MTEB~\citep{ciancone2024mtebfrenchresourcesfrenchsentence} benchmark (26 datasets in the French subset). Please refer to Table~\ref{table:full_fr_pl_c_mteb} for the scores of bge-multilingual-gemma2 per dataset.}
\vspace{10pt}
\label{table:fr_mteb_results}
\end{table}

\begin{table}[!t]
    \centering
    \resizebox{0.9\textwidth}{!}{
        \begin{tabular}{l|ccccc|c}
        \hline
        Task & Retr. & Clust. & PairClass. & Class. & STS & Avg. \\  
        \# of datasets $\rightarrow$ & 11 & 1 & 4 & 7 & 3 & 26 \\ \hline
        gte-multilingual-base & 46.40 & 33.67 & 85.45 & 60.15 & 68.92 & 58.22 \\
        multilingual-e5-large & 48.98 & 33.88 & 85.50 & 63.82 & 66.91 & 60.08 \\
        mmlw-roberta-large & 52.71 & 31.16 & 89.13 & 66.39 & 70.59 & 63.23 \\
        gte-Qwen2-1.5B-instruct & 51.88 & 44.59 & 84.87 & 72.29 & 68.12 & 64.04 \\
        gte-Qwen2-7B-instruct & 54.69 & \textbf{51.36} & 88.48 & 77.84 & \textbf{70.86} & 67.86 \\
        \textbf{bge-multilingual-gemma2} & \textbf{59.41} & 50.29 & \textbf{89.62} & \textbf{77.99} & 70.64 & \textbf{70.00}  \\
        \hline
        \end{tabular}
    }
\vspace{-5pt}
\caption{Results on the PL-MTEB~\citep{poswiata2024plmteb} benchmark (26 datasets in the Polish subset). Please refer to Table~\ref{table:full_fr_pl_c_mteb} for the scores of bge-multilingual-gemma2 per dataset.}
\vspace{10pt}
\label{table:pl_mteb_results}
\end{table}

\begin{table}[!h]
    \centering
    \resizebox{\textwidth}{!}{
        \begin{tabular}{l|cccccc|c}
        \hline
        Task & Retr. & Rerank. & Clust. & PairClass. & Class. & STS & Avg. \\  
        \# of datasets $\rightarrow$ & 8 & 4 & 4 & 2 & 9 & 8 & 35 \\ \hline
        multilingual-e5-large & 63.66 & 56.00 & 48.23 & 69.89 & 67.34 & 48.29 & 58.81 \\
        e5-mistral-7b-instruct & 61.75 & 61.86 & 52.30 & 72.19 & 70.17 & 50.22 & 60.81 \\
        gte-multilingual-base & 71.95 & 68.17 & 47.48 & 78.34 & 64.27 & 52.73 & 62.72 \\
        bge-large-zh-v1.5 & 70.46 & 65.84 & 48.99 & 81.6 & 69.13 & 56.25 & 64.53 \\
        gte-Qwen2-1.5B-instruct & 71.86 & 68.21 & 54.61 & 86.91 & 71.12 & 60.96 & 67.65 \\
        gte-Qwen2-7B-instruct & \textbf{76.03} & \textbf{68.92} & \textbf{66.06} & \textbf{87.48} & \textbf{75.09} & \textbf{65.33} & \textbf{72.05} \\
        \textbf{bge-multilingual-gemma2} & 73.73 & 68.28 & 59.3 & 86.67 & 74.11 & 56.87 & 68.44 \\
        \hline
        \end{tabular}
    }
\vspace{-5pt}
\caption{Results on the C-MTEB~\citep{xiao2024c} benchmark (35 datasets in the Chinese subset). Please refer to Table~\ref{table:full_fr_pl_c_mteb} for the scores of bge-multilingual-gemma2 per dataset.}
\vspace{10pt}
\label{table:c_mteb_results}
\end{table}

\textbf{MTEB}. The evaluation results of bge-multilingual-gemma2 on the MTEB benchmark are available in Table~\ref{table:mteb}. The detailed results for each dataset are available in Table~\ref{table:fullmteb}. We can also observe that bge-multilingual-gemma2 achieves good performance on the MTEB benchmark.

\textbf{AIR-Bench}. For the QA task in AIR-Bench, we perform evaluation on all of the 13 datasets in 24.04 version, which consists of 8 English datasets and 5 Chinese datasets. For the Long-Doc task in AIR-Bench, we perform evaluation on all of the 15 datasets in 24.04 version, which are all English datasets. As shown in Table~\ref{table:air_qa_en_zh} and Table~\ref{table:air_longdoc}, bge-multilingual-gemma2 also achieves excellent performance in the out-of-distribution (OOD) evaluation on AIR-Bench, which indicates that our model has excellent generalization ability.

\begin{table}[!t]
\centering
\begin{small}
\setlength{\tabcolsep}{4pt}
\begin{tabular}{l|cccccccc|c}
\hline
Domain & wiki & web & news & healthcare & law & finance & arxiv & msmarco & Avg. \\
\# of datasets $\rightarrow$ & 2 & 2 & 2 & 2 & 1 & 2 & 1 & 1 & 13 \\ \hline
bge-m3 (Dense) & 61.42 & 48.86 & 44.40 & 45.74 & \textbf{26.68} & 41.85 & 40.76 & 54.40 & 46.65 \\
multilingual-e5-large & 57.16 & 42.91 & 41.61 & 42.18 & 19.66 & 37.38 & 36.93 & 54.44 & 42.58 \\
e5-mistral-7b-instruct & 58.82 & 45.18 & 42.08 & 46.06 & 19.32 & 40.45 & \textbf{44.78} & 59.03 & 45.26 \\
gte-Qwen2-1.5B-instruct & 55.04 & 42.95 & 37.30 & 44.50 & 11.95 & 40.24 & 32.06 & 49.74 & 41.06 \\
gte-Qwen2-7B-instruct & 64.95 & 51.59 & \textbf{48.55} & \textbf{46.51} & 22.31 & \textbf{42.42} & 40.27 & 58.39 & \textbf{48.38} \\
\textbf{bge-multilingual-gemma2} & \textbf{65.50} & \textbf{51.81} & 47.46 & 44.68 & 22.58 & 40.45 & 23.28 & \textbf{63.14} & 46.83 \\
\hline
\end{tabular}
\end{small}
\vspace{-5pt}
\caption{QA (en \& zh, nDCG@10) performance on AIR-Bench 24.04.}
\vspace{10pt}
\label{table:air_qa_en_zh}
\end{table}

\begin{table}[h]
\centering
\setlength{\extrarowheight}{1.5pt}
\resizebox{\textwidth}{!}{
% \scalebox{0.9}{
    \begin{tabular}{ll}
    \toprule
    Task Name & Instruction Template \\ \midrule \midrule
    C-MTEB & \\ \midrule
    CLSClusteringS2S & Identify the main category of scholar papers based on the titles. \\
    CLSClusteringP2P & Identify the main category of scholar papers based on the titles and abstracts. \\
    ThuNewsClusteringS2S & Identify the topic or theme of the given news articles based on the titles. \\
    ThuNewsClusteringP2P & Identify the topic or theme of the given news articles based on the titles and contents. \\
    T2Reranking & Given a Chinese search query, retrieve web passages that answer the question. \\
    MMarcoReranking & Given a Chinese search query, retrieve web passages that answer the question. \\
    CMedQAv1 & Given a Chinese community medical question, retrieve replies that best answer the question. \\
    CMedQAv2 & Given a Chinese community medical question, retrieve replies that best answer the question. \\
    Ocnli & Retrieve semantically similar text. \\
    Cmnli & Retrieve semantically similar text. \\
    T2Retrieval & Given a Chinese search query, retrieve web passages that answer the question. \\
    MMarcoRetrieval & Given a web search query, retrieve relevant passages that answer the query. \\
    DuRetrieval & Given a Chinese search query, retrieve web passages that answer the question. \\
    CovidRetrieval & Given a question on COVID-19, retrieve news articles that answer the question. \\
    CmedqaRetrieval & Given a Chinese community medical question, retrieve replies that best answer the question. \\
    EcomRetrieval & Given a user query from an e-commerce website, retrieve description sentences of relevant products. \\
    MedicalRetrieval & Given a medical question, retrieve user replies that best answer the question. \\
    VideoRetrieval & Given a video search query, retrieve the titles of relevant videos. \\ \midrule 
    PL-MTEB & \\ \midrule
    CBD & Classify the sentiment of polish tweet reviews. \\
    PolEmo2.0-IN & Classify the sentiment of in-domain (medicine and hotels) online reviews. \\
    PolEmo2.0-OUT & Classify the sentiment of out-of-domain (products and school) online reviews. \\
    AllegroReviews & Classify the sentiment of reviews from e-commerce marketplace Allegro. \\
    PAC & \begin{tabular}[c]{@{}l@{}}Classify the sentence into one of the two types: \\
    ``BEZPIECZNE\_POSTANOWIENIE\_UMOWNE'' and ``KLAUZULA\_ABUZYWNA''. \end{tabular} \\
    8TagsClustering & \begin{tabular}[c]{@{}l@{}}Identify of headlines from social media posts in Polish  into 8 categories: \\  film, history, food, medicine, motorization, work, sport and technology. \end{tabular}  \\
    SICK-E-PL & Retrieve semantically similar text. \\
    PPC & Retrieve semantically similar text. \\
    CDSC-E & Retrieve semantically similar text. \\
    PSC & Retrieve semantically similar text. \\
    ArguAna-PL    &     Given a claim, find documents that refute the claim. \\
    DBPedia-PL    &     Given a query, retrieve relevant entity descriptions from DBPedia. \\
    FiQA-PL    &     Given a financial question, retrieve user replies that best answer the question. \\
    HotpotQA-PL    &     Given a multi-hop question, retrieve documents that can help answer the question. \\
    MSMARCO-PL    &     Given a web search query, retrieve relevant passages that answer the query. \\
    NFCorpus-PL    &     Given a question, retrieve relevant documents that best answer the question. \\
    NQ-PL    &     Given a question, retrieve Wikipedia passages that answer the question. \\
    Quora-PL    &     Given a question, retrieve questions that are semantically equivalent to the given question. \\
    SCIDOCS-PL    &     Given a scientific paper title, retrieve paper abstracts that are cited by the given paper. \\
    SciFact-PL    &     Given a scientific claim, retrieve documents that support or refute the claim. \\
    Touche2020    &     Given a question, retrieve detailed and persuasive arguments that answer the question. \\
    TRECCOVID-PL    &     Given a query, retrieve documents that answer the query. \\ \midrule
    FR-MTEB & \\ \midrule
    MasakhaNEWSClassification & \begin{tabular}[c]{@{}l@{}}Classify the given news article into one of the seven topic categories:\\  politics, sports, health, business, entertainment, technology, and religion. \end{tabular}  \\
    AlloProfClusteringP2P & Identify the main category of Allo Prof document based on the titles and descriptions. \\
    AlloProfClusteringS2S & Identify the main category of Allo Prof document based on the titles. \\
    HALClusteringS2S & Identify the main category of academic passage based on the titles and contents. \\
    MasakhaNEWSClusteringP2P & Identify the topic or theme of the given news articles based on the titles and contents. \\
    MasakhaNEWSClusteringS2S & Identify the topic or theme of the given news articles based on the titles. \\
    MLSUMClusteringP2P & Identify the topic or theme of the given articles based on the titles and contents. \\
    MLSUMClusteringS2S & Identify the topic or theme of the given articles based on the titles. \\
    AlloprofReranking & Given a question, retrieve passages that answer the question. \\
    OpusparcusPC & Retrieve semantically similar text. \\
    PawsXPairClassification & Retrieve semantically similar text. \\
    SyntecReranking & Given a question, retrieve passages that answer the question. \\
    AlloprofRetrieval & Given a question, retrieve passages that answer the question. \\
    BSARDRetrieval & Given a question, retrieve passages that answer the question. \\
    SyntecRetrieval & Given a question, retrieve passages that answer the question. \\
    XPQARetrieval & Given a question, retrieve passages that answer the question. \\
    MintakaRetrieval & Given a question, retrieve passages that answer the question. \\
    \bottomrule
    \end{tabular}
}
\caption{The additional instruction we used on the C-MTEB, PL-MTEB and FR-MTEB benchmarks. These instructions are adopted from gte-Qwen2-7B-instruct~\citep{li2023towards}. To ensure sentence completeness, we add a period at the end.}
\label{table:multilingual_evalinstructions}
\end{table}

\begin{table}[h]
\begin{center}
\setlength{\extrarowheight}{2.5pt}
\begin{small}\begin{tabular}{l|c||l|c}
\hline
% \begin{tabular}[c]{@{}c@{}} bge-multilingual-\\gemma2 \end{tabular}
Dataset & Result & Dataset & Result \\ \hline \hline
\multicolumn{2}{l||}{FR-MTEB} & \multicolumn{2}{l}{PL-MTEB} \\ \hline
AlloprofRetrieval & 58.50 & ArguAna-PL & 59.71 \\
BSARDRetrieval & 28.52 & DBPedia-PL & 43.19 \\
MintakaRetrieval (fr) & 62.53 & FiQA-PL & 46.12 \\
SyntecRetrieval & 90.37 & HotpotQA-PL & 77.03 \\
XPQARetrieval (fr) & 77.42 & MSMARCO-PL & 72.69 \\
AlloprofReranking & 78.62 & NFCorpus-PL & 36.72 \\
SyntecReranking & 91.83 & NQ-PL & 56.85 \\
AlloProfClusteringP2P & 71.20 & Quora-PL & 84.47 \\
AlloProfClusteringS2S & 59.64 & SCIDOCS-PL & 19.53 \\
HALClusteringS2S & 28.19 & SciFact-PL & 74.43 \\
MLSUMClusteringP2P (fr) & 47.75 & TRECCOVID-PL & 82.75 \\
MLSUMClusteringS2S (fr) & 47.46 & 8TagsClustering & 50.29 \\
MasakhaNEWSClusteringP2P (fra) & 73.86 & CDSC-E & 78.23 \\
MasakhaNEWSClusteringS2S (fra) & 67.24 & PPC & 95.43 \\
OpusparcusPC (fr) & 100.00 & PSC & 99.24 \\
PawsXPairClassification (fr) & 70.14 & SICK-E-PL & 85.58 \\
AmazonReviewsClassification (fr) & 55.19 & AllegroReviews & 65.00 \\
MasakhaNEWSClassification (fra) & 82.49 & CBD & 84.13 \\
MassiveIntentClassification (fr) & 79.60 & MassiveIntentClassification (pl) & 79.41 \\
MassiveScenarioClassification (fr) & 82.18 & MassiveScenarioClassification (pl) & 81.93 \\
MTOPDomainClassification (fr) & 97.20 & PAC & 67.24 \\
MTOPIntentClassification (fr) & 93.07 & PolEmo2.0-IN & 90.42 \\
STS22 (fr) & 83.28 & PolEmo2.0-OUT & 77.77 \\
STSBenchmarkMultilingualSTS (fr) & 85.09 & CDSC-R & 90.97 \\
SICKFr & 79.39 & SICK-R-PL & 78.16 \\
SummEvalFr & 31.26 & STS22 (pl) & 42.79 \\ \hline
\textbf{FR-MTEB Average (23)} & \textbf{70.08} & \textbf{PL-MTEB Average (23)} & \textbf{70.00} \\ \hline \hline
\multicolumn{4}{l}{C-MTEB} \\ \hline
CmedqaRetrieval & 42.21 & AmazonReviewsClassification (zh)  & 54.34 \\
CovidRetrieval & 77.46 & IFlyTek & 49.94 \\
DuRetrieval & 90.46 & JDReview & 88.91 \\
EcomRetrieval & 69.3 & MassiveIntentClassification (zh-CN)  & 78.19 \\
MedicalRetrieval & 62.02 & MassiveScenarioClassification (zh-CN) & 82.58 \\
MMarcoRetrieval & 84.7 & MultilingualSentiment & 78.91 \\
T2Retrieval & 86.26 & OnlineShopping & 94.59 \\
VideoRetrieval & 77.4 & TNews & 50.26 \\
CMedQAv1 & 84.62 & Waimai & 89.26 \\
CMedQAv2 & 85.60 & AFQMC  & 47.17 \\
MMarcoReranking & 35.43 & ATEC & 50.75 \\
T2Reranking & 67.48 & BQ & 62.02 \\
CLSClusteringP2P & 54.65 & LCQMC & 75.95 \\
CLSClusteringS2S & 63.68 & PAWSX & 30.57 \\
ThuNewsClusteringP2P & 64.32 & QBQTC & 38.98 \\
ThuNewsClusteringS2S & 54.57 & STS22 (zh) & 68.68 \\
Cmnli & 90.13 & STSB & 80.87 \\ \cline{3-4}
Ocnli & 83.21 & \textbf{C-MTEB Average (35)} & \textbf{68.44} \\
\hline
\end{tabular}
\end{small}
\end{center}
\vspace{-0.4cm}
\caption{Results of \textbf{bge-multilingual-gemma2} for each dataset in the FR-MTEB, PL-MTEB and C-MTEB benchmarks.}
\label{table:full_fr_pl_c_mteb}
\end{table}

\clearpage
\newpage
\section{Lightweight Re-ranker}
\label{reranker}

We have also introduced a lightweight version of the reranker, which incorporates both depth and width compression techniques. Specifically, depth compression is implemented on a layerwise method, allowing for the selective adjustment of the number of layers according to the desired output. Regarding width compression, it is configured to execute token compression at predetermined layers, whereby $n$ tokens are merged into a single token.

For the input template, we use the following format:
\begin{equation}
    \text{A:} \hspace{0.2cm} \{\text{query}\} \newline
    \hspace{0.2cm} \text{B:} \hspace{0.2cm} \{\text{passage}\} \newline
    \hspace{0.2cm} \{\text{prompt}\}
\end{equation}
where the prompt inquires about the relationship between A and B, e.g., \textit{Predict whether passage B contains an answer to query A.} And we use the logits of \textit{Yes} as our reranking score.

Considering the depth compression generates output scores at each layer, we extract the linear layer connected to the logits for the \textit{Yes} prediction in the language model head. This extracted linear layer is then appended to each layer, allowing every layer to compute a reranking score.

\subsection{Setup}

\textbf{LLM}. Our objective is to develop a multilingual version of the lightweight reranker. Considering the extensive vocabulary necessitated by multilingual support, we employ Gemma-2-9b \citep{team2024gemma} as the backbone for our reranker.

\textbf{Dataset}. We evaluate the performance of our reranker \textit{bge-reranker-v2.5-gemma-lightweight} on BEIR \citep{thakur2021beir} and MIRACL \citep{zhang2023miracl}. The BEIR benchmark encompasses a variety of text retrieval tasks across multiple domains, while MIRACL serves as a significant dataset for multilingual evaluation, featuring 18 distinct languages.

\textbf{Training Data}. To enhance the multilingual capabilities and retrieval performance of the Reranker, we utilize the BGE-M3 dataset \citep{chen2024bge}, along with Arguana, HotpotQA, and FEVER, for the training process.

\textbf{Training Detail}. The reranker is trained using contrastive loss. Furthermore, LoRA is employed for fine-tuning, where the LoRA rank is set to 64 and the LoRA alpha is set to 32, accompanied by a learning rate of 1e-4. During the training process, a batch size of 128 is utilized, and 15 hard negatives are assigned to each query. At the same time, the training of the reranker employs self-distillation, wherein the final layer serves as the teacher for preceding layers. Throughout this training process, KL divergence loss is utilized. During training, we randomly select a width compression strategy and train all depth compression strategies. Regarding depth compression, we support outputs from 8 to 42 layers. Regarding width compression, we support compression ratios of 1, 2, 4, and 8, and support width compression at 8, 16, 24, 32, and 40 layers. During the training process, we utilized four types of prompts: query to passage, query to query, passage to passage, and argument to counter-argument. The specific application of these prompts was dependent on the type of dataset used, as shown in Table \ref{table:reins}.

\textbf{Evaluation}. On the BEIR benchmark, we rerank the top-100 retrieval results of bge-large-en-v1.5 and E5-mistral-7b-instruct. On the MIRACL dataset, we rerank the top-100 retrieval results of bge-m3 (dense). The instructions for evaluation are shown in Table \ref{table:rerank_inst}.

\begin{table}[h]
\begin{center}
\begin{small}
\setlength{\tabcolsep}{4.5pt}
\setlength{\extrarowheight}{2pt}
\resizebox{\textwidth}{!}{
\begin{tabular}{ll}
\toprule
Task Type & Instruction Template \\ \midrule
query to passage & Predict whether passage B contains an answer to query A. \\
query to query & Predict whether queries A and B are asking the same thing. \\
passage to passage & Predict whether passages A and B have the same meaning. \\
argument to counter-argument & Predict whether argument A and counterargument B express contradictory opinions. \\ \bottomrule
\end{tabular}
}
\end{small}
\end{center}
\caption{The training instructions we used for reranker.}
\label{table:reins}
\end{table}

\begin{table}[t]
\begin{center}
\begin{small}
\setlength{\tabcolsep}{4.5pt}
\setlength{\extrarowheight}{2pt}
\resizebox{\textwidth}{!}{
\begin{tabular}{ll}
\toprule
Task Name & Instruction Template \\ \midrule
ArguAna & Predict whether argument A and counterargument B express contradictory opinions. \\
ClimateFEVER & Predict whether passage B contains an answer to query A. \\
CQADupstack & Predict whether queries A and B are asking the same thing. \\
DBPedia & Predict whether passage B contains an answer to query A. \\
FEVER & Predict whether passage B contains an answer to query A. \\
FiQA2018 & Predict whether passage B contains an answer to query A. \\
HotpotQA & Predict whether passage B contains an answer to query A. \\
MSMARCO & Predict whether passage B contains an answer to query A. \\
NFCorpus & Predict whether passage B contains an answer to query A. \\
Natural Question & Predict whether passage B contains an answer to query A. \\
QuoraRetrieval & Predict whether queries A and B are asking the same thing. \\
SCIDOCS & Predict whether passage B contains an answer to query A. \\
SciFact & Predict whether passage B contains an answer to query A. \\
Touche2020 & Predict whether passage B contains an answer to query A. \\
TREC-COVID & Predict whether passage B contains an answer to query A. \\ \midrule
MIRACL& Predict whether passage B contains an answer to query A. \\ \bottomrule
\end{tabular}
}
\end{small}
\end{center}
\caption{The instructions we used for the BEIR benchmark and MIRACL dataset for reranker.}
\label{table:rerank_inst}
\end{table}

\subsection{Main Results}

\begin{table}[h]
\hspace*{-1.6cm}
\begin{small}
\centering
\begin{tabular}{c|c|ccccc}
\hline
BEIR & \begin{tabular}[c]{@{}l@{}}bge-large-\\ en-v1.5\end{tabular} & \begin{tabular}[c]{@{}l@{}}bge-rerank\\ er-v2-m3\end{tabular} & \begin{tabular}[c]{@{}l@{}}jina-reranker-v2-\\ base-multilingual\end{tabular} & \begin{tabular}[c]{@{}l@{}}bge-reranker-\\ v2-gemma\end{tabular} & \begin{tabular}[c]{@{}l@{}}bge-reranker-v2.5-\\ gemma-lightweight\end{tabular} & \begin{tabular}[c]{@{}l@{}}bge-reranker-v2.5-\\ gemma-lightweight\end{tabular} \\ \hline
Save Flops & - & - & - & - & 60\% & 0 \\ \hline
ArguAna & 63.54 & 37.70 & 52.23 & 78.68 & 86.04 & 86.16 \\
ClimateFEVER & 36.49 & 37.99 & 34.65 & 39.07 & 48.41 & 48.48 \\
CQA & 42.23 & 38.24 & 40.21 & 45.85 & 49.18 & 48.9 \\
DBPedia & 44.16 & 48.15 & 49.31 & 49.92 & 51.98 & 52.11 \\
FEVER & 87.17 & 90.15 & 92.44 & 90.15 & 94.71 & 94.69 \\
FiQA2018 & 44.97 & 49.32 & 45.88 & 49.32 & 60.48 & 60.95 \\
HotpotQA & 74.11 & 84.51 & 81.81 & 86.15 & 87.84 & 87.89 \\
MSMARCO & 42.48 & 47.79 & 47.83 & 48.07 & 47.23 & 47.26 \\
NFCorpus & 38.12 & 34.85 & 37.73 & 39.73 & 41.40 & 41.64 \\
NQ & 55.04 & 69.37 & 67.35 & 72.60 & 75.37 & 75.58 \\
QuoraRetrieval & 89.06 & 89.13 & 87.81 & 90.37 & 91.25 & 91.18 \\
SCIDOCS & 22.62 & 18.25 & 20.21 & 21.65 & 23.71 & 23.87 \\
SciFact & 74.64 & 73.08 & 76.93 & 77.22 & 80.5 & 80.38 \\
Touche2020 & 25.08 & 35.68 & 32.45 & 35.68 & 30.64 & 31.09 \\
TRECCOVID & 74.89 & 83.39 & 80.89 & 85.51 & 84.26 & 84.85 \\ \hline
\textbf{Mean} & 54.31 & 55.36 & 56.52 & 60.71 & 63.10 & \textbf{63.67} \\ \hline
\end{tabular}
\end{small}
\caption{The performance of various rerankers on BEIR benchmark (based on bge-large-en-v1.5).}
\label{table:beir_rerank_bge}
\end{table}

\begin{table}[!t]
\begin{small}
\centering
\begin{tabular}{c|c|ccc}
\hline
BEIR & \begin{tabular}[c]{@{}l@{}}E5-mistral-\\ 7b-instruct\end{tabular} & \begin{tabular}[c]{@{}l@{}}bge-reranker-\\ v2-gemma\end{tabular} & \begin{tabular}[c]{@{}l@{}}bge-reranker-v2.5-\\ gemma-lightweight\end{tabular} & \begin{tabular}[c]{@{}l@{}}bge-reranker-v2.5-\\ gemma-lightweight\end{tabular} \\ \hline
Save Flops & - & - & 60\% & 0 \\ \hline
ArguAna & 61.80 & 79.05 & 86.02 & 86.58 \\
ClimateFEVER & 38.37 & 37.66 & 47.27 & 47.13 \\
CQA & 42.97 & 46.16 & 49.06 & 49.53 \\
DBPedia & 48.84 & 50.77 & 52.45 & 52.87 \\
FEVER & 87.82 & 91.36 & 94.85 & 95.19 \\
FiQA2018 & 56.58 & 50.96 & 58.81 & 61.19 \\
HotpotQA & 75.72 & 86.99 & 88.49 & 88.82 \\
MSMARCO & 43.06 & 48.35 & 47.65 & 47.40 \\
NFCorpus & 38.58 & 39.25 & 42.28 & 42.17 \\
NQ & 63.56 & 73.44 & 75.00 & 76.28 \\
QuoraRetrieval & 89.59 & 90.44 & 91.09 & 91.18 \\
SCIDOCS & 16.30 & 20.77 & 22.20 & 22.69 \\
SciFact & 76.26 & 77.78 & 79.94 & 80.98 \\
Touche2020 & 26.24 & 35.79 & 28.69 & 31.17 \\
TRECCOVID & 87.07 & 88.13 & 86.61 & 87.36 \\ \hline
\textbf{Mean} & 56.85 & 61.13 & 63.36 & \textbf{64.04} \\ \hline
\end{tabular}
\end{small}
\caption{The performance of various rerankers on BEIR benchmark (based on E5-mistral-7b-insturct).}
\label{table:beir_rerank_e5}
\end{table}

\textbf{BEIR}.
We rerank the retrieval results from the BEIR dataset using two models, bge-large-en-v1.5 and E5-Mistral-7b-Instruct, and we rerank top-100 retrieval results from these models. We conduct both a comprehensive evaluation and a lightweight evaluation. In the lightweight evaluation, we select a compression ratio of 2, a width compression factor of 8, and a depth of the output layer set to 25. This configuration results in a 60\% FLOPs.

Tables \ref{table:beir_rerank_bge} and \ref{table:beir_rerank_e5} present the evaluation results for the BEIR benchmark. It indicates that bge-reranker-v2.5-gemma2-lightweight records exceptional performance in enhancing both bge-large-en-v1.5 and E5-Mistral-7b-Instruct retrieval outcomes. Furthermore, there exists a positive correlation between the initial retrieval quality and the subsequent reranking performance, when  reranking the retrieval results from E5-Mistral-Instruct, our reranker achieves improved performance. Additionally, the implementation of the lightweight model variant results in only a marginal decline in performance while achieving a significant reduction in FLOPs. 

\begin{table}[h]
\hspace*{-1.6cm}
\begin{small}
\centering
\begin{tabular}{c|c|ccccc}
\hline
Language & bge-m3 (Dense) & \begin{tabular}[c]{@{}l@{}}bge-reranker-\\ v2-gemma\end{tabular} & \begin{tabular}[c]{@{}l@{}}bge-rerank\\ er-v2-m3\end{tabular} & \begin{tabular}[c]{@{}l@{}}bge-reranker-\\ v2-gemma\end{tabular} & \begin{tabular}[c]{@{}l@{}}bge-reranker-v2.5-\\ gemma-lightweight\end{tabular} & \begin{tabular}[c]{@{}l@{}}bge-reranker-v2.5-\\ gemma-lightweight\end{tabular} \\
\hline
FLOPS & - & - & - & - & 60\% & 0 \\ \hline
ar & 78.4 & 73.4 & 81.7 & 82.3 & 82.5 & 82.8 \\
bn & 80.0 & 81.9 & 84.6 & 85.0 & 87.8 & 87.6 \\
en & 56.9 & 58.9 & 63.5 & 66.6 & 68.6 & 69.3 \\
es & 56.1 & 58.6 & 64.4 & 65.3 & 67.6 & 67.8 \\
fa & 60.9 & 60.5 & 65.7 & 65.5 & 67.5 & 67.4 \\
fi & 78.6 & 77.2 & 82.4 & 82.6 & 82.8 & 83.3 \\
fr & 58.3 & 56.1 & 63.7 & 65.4 & 68.5 & 68.5 \\
hi & 59.5 & 62.7 & 68.5 & 69.4 & 71.4 & 71.3 \\
id & 56.1 & 59.6 & 62.7 & 61.2 & 63.8 & 63.8 \\
ja & 72.8 & 72.7 & 80.0 & 79.7 & 82.8 & 83.6 \\
ko & 69.9 & 74.0 & 73.8 & 75.1 & 75.9 & 75.7 \\
ru & 70.1 & 67.1 & 76.9 & 78.3 & 79.8 & 80.1 \\
sw & 78.7 & 78.1 & 82.3 & 81.8 & 84.8 & 85.1 \\
te & 86.2 & 85.8 & 89.4 & 89.6 & 90.8 & 90.8 \\
th & 82.6 & 81.2 & 85.3 & 86.1 & 88.1 & 88.7 \\
zh & 62.7 & 63.0 & 65.2 & 66.8 & 69.9 & 69.9 \\
de & 56.7 & 58.2 & 62.7 & 64.0 & 65.8 & 65.6 \\
yo & 81.8 & 84.2 & 87.4 & 85.9 & 89.6 & 89.8 \\ \hline
Mean (18) & 69.2 & 69.6 & 74.4 & 75.0 & 77.1 & \textbf{77.3} \\
\hline
\end{tabular}
\end{small}
\caption{Comparison of MIRACL dev nDCG@10 scores across various rerankers (based on bge-m3 (Dense)).}
\label{table:miracl_rerank}
\end{table}

\textbf{MIRACL}.
% To evaluate the multilingual capabilities of the reranker, we conduct an assessment using the MIRACL dataset. The results of this evaluation are presented in Table \ref{table:miracl_rerank}. We rerank the top 100 retrieval results of the bge-m3 (dense) model, achieving excellent performance across each individual dataset as well as on average. Additionally, the impact of model compression on multilingual performance was found to be minimal, with the lightweight model maintaining nearly the same effectiveness as the original.
We further evaluate the multilingual capabilities of the reranker using the MIRACL dataset, with the results presented in Table \ref{table:miracl_rerank}. The reranking is conducted based on the top 100 retrieval results obtained from the bge-m3 (dense) model. The reranker demonstrates a significant improvement in retrieval accuracy across each dataset and outperforms other multilingual rerankers. Notably, compared to monolingual (English) retrieval, the multilingual retrieval experienced minimal negative effects from model lightweighting, essentially maintaining the original performance of the model.

\end{document}